\documentclass[11pt,a4paper]{article}
\usepackage{jheppubedit}
\usepackage{enumitem}
\usepackage{amsmath}
\usepackage{amsfonts}
\usepackage{graphicx}
\usepackage{caption}

\title{Perturbative String Thermodynamics near Black Hole Horizons}

\author[a]{Thomas G. Mertens,}
\author[a]{Henri Verschelde}
\author[b]{and Valentin I. Zakharov}
\affiliation[a]{Ghent University, Department of Physics and Astronomy\\
Krijgslaan, 281-S9, 9000 Gent, Belgium}
\affiliation[b]{ITEP, B. Cheremushkinskaya 25, Moscow, 117218 Russia,\\
Moscow Inst Phys \& Technol, Dolgoprudny, Moscow Region, 141700 Russia ,\\
School of Biomedicine, Far Eastern Federal University, Sukhanova str 8, 
Vladivostok 690950 Russia
}
\emailAdd{thomas.mertens@ugent.be}
\emailAdd{henri.verschelde@ugent.be}
\emailAdd{vzakharov@itep.ru}
\abstract{We provide further computations and ideas to the problem of near-Hagedorn string thermodynamics near (uncharged) black hole horizons, building upon our earlier work \cite{Mertens:2013zya}. The relevance of long strings to one-loop black hole thermodynamics is emphasized. We then provide an argument in favor of the absence of $\alpha'$-corrections for the (quadratic) heterotic thermal scalar action in Rindler space. We also compute the large $k$ limit of the cigar orbifold partition functions (for both bosonic and type II superstrings) which allows a better comparison between the flat cones and the cigar cones. A discussion is made on the general McClain-Roth-O'Brien-Tan theorem and on the fact that different torus embeddings lead to different aspects of string thermodynamics. The black hole/string correspondence principle for the 2d black hole is discussed in terms of the thermal scalar. Finally, we present an argument to deal with arbitrary higher genus partition functions, suggesting the breakdown of string perturbation theory (in $g_s$) to compute thermodynamical quantities in black hole spacetimes.}

\begin{document}

\maketitle

\section{Introduction}

String theory in a black hole background is reasonably understood for extremal black holes (where the microstates can be identified as perturbative string states, solitons and branes within string theory).\footnote{See e.g. \cite{Strominger:1996sh} for the earliest account of this. The vast amount of literature that follows is too numerous to be cited here.} Non-extremal black holes on the other hand are (in spite of several influential ideas) still largely understood on a qualitative level only. The main line of thought that pervades the literature on this topic is the idea that a long string(s) should give the necessary microstructure to such black holes.\footnote{A recent interesting paper \cite{Martinec:2014gka} has some suggestive ideas concerning the precise way in which the microstructure is accounted for by long strings.} In flat space, it was argued for many years ago that the near-Hagedorn string gas should behave as a few long strings whose spatial form describes random walks in the ambient space \cite{Atick:1988si}\cite{Mitchell:1987hr}\cite{Mitchell:1987th}\cite{Bowick:1989us}\cite{Deo:1989bv}\cite{Horowitz:1997jc}\cite{Barbon:2004dd}. For the black hole case, this long string should also account for the black hole membrane where (following some early ideas) the black hole degrees of freedom are stored. The picture that emerges is that a (uncharged) black hole is surrounded by a long string at string length distance from the event horizon. The fact that strings tend to elongate when approaching the horizon was shown from a single string perspective in \cite{Susskind:1993ws}\cite{Susskind:1993ki}\cite{Susskind:1994uu}\cite{Susskind:1993aa}. The study of the canonical ensemble of a string gas surrounding a black hole horizon was also started around the same period \cite{Dabholkar:1994ai}\cite{Lowe:1994ah}. In \cite{Mertens:2013zya} we combined thermal scalar field theory computations with the explicit random walk picture of the long string \cite{Kruczenski:2005pj}\cite{theory} to provide a realization of Susskind's picture in the canonical ensemble. We arrived at the conclusion that the Hawking temperature equals the Hagedorn temperature and hence in this sense the long string phase prevails. \\

\noindent The main goal of this paper is to further analyze some of the more puzzling parts of this story. \\

\noindent The paper is organized as follows. Section \ref{recap} contains a review of the results presented in \cite{Mertens:2013zya} in which we will provide more details on the puzzles that were left open in that work. Then, as an appetizer, in section \ref{relevance} we discuss the role and relevance of long strings in one-loop black hole thermodynamics. In particular we argue that a truncation to the one-loop thermodynamics of only the massless modes is incapable of approximating the one-loop string result. Then in section \ref{heterotic}, we discuss Rindler space for the heterotic string, a problem that was left open in \cite{Mertens:2013zya}. We noted there that it appears that the thermal scalar action for the heterotic string in Euclidean Rindler space without any $\alpha'$-corrections is capable of reproducing expected results. Here we present an argument in favor of this. In section \ref{flatlimit}, we present a detailed comparison of the large $k$ limit of the bosonic and type II superstring cigar orbifold partition functions. This generalizes the comparison done in \cite{Mertens:2013zya}. Next, section \ref{modDom} provides some ideas on how the different approaches to string thermodynamics fit together with a special emphasis on the difference between the modular strip and the fundamental domain. We also discuss subtleties on torus embeddings that are expected to correspond to the Susskind-Uglum interactions on the horizon \cite{Susskind:1994uu}. After that, in section \ref{remarks}, we make some comments on the string/black hole transition in the $SL(2,\mathbb{R})/U(1)$ cigar CFT from a thermal scalar perspective. Finally, in section \ref{higher}, we present an analysis of higher genus corrections (closely following \cite{Brigante:2007jv}). The crux of the matter is that it appears that the entire genus expansion should be resummed, even though for each fixed genus a random walk picture (with self-intersections) emerges.\\

\noindent Appendix \ref{cigspectr} contains the cigar CFT string spectrum. Several technical calculations on the McClain-Roth-O'Brien-Tan theorem in curved spacetime, the cigar CFT and a discussion on normalizable and non-normalizable operators are given in the remaining appendices.

\section{Recap of the thermal scalar near black holes}
\label{recap}
High temperature string theory is known to exhibit divergences in its one-loop thermodynamical quantities due to the enormous degeneracy of high-energy states. This critical temperature is called the Hagedorn temperature. It can be related to a string state on the thermal manifold (Wick rotated background with periodically identified temporal dimension) that becomes massless precisely at this temperature: this state is singly wound ($w=\pm1$) around the thermal direction and is called the thermal scalar. \\

\noindent In \cite{Mertens:2013zya} we set out to study this question for black hole geometries where the thermal circle is cigar-like and pinches off at the horizon. We were interested in finding the thermal spectrum on Euclidean Rindler space (the near-horizon approximation to a large class of (uncharged) black holes):
\begin{equation}
\label{snorm}
ds^2 = \left(\frac{\rho^2}{\alpha'}\right)d\tau^2 + d\rho^2 + d\mathbf{x}^2_{\perp}.
\end{equation}
In this geometry, one identifies $\tau \sim \tau + 2\pi\sqrt{\alpha'}$ to avoid a conical singularity at the origin (horizon of the black hole); we call this (inverse) temperature $\beta_R$.
To find the thermal spectrum, in \cite{Mertens:2013zya} we have followed the strategy proposed by \cite{Giveon:2012kp}\cite{Giveon:2013ica}\cite{Giveon:2014hfa} to take the small curvature (large $k$) limit of the $SL(2,\mathbb{R})/U(1)$ cigar CFT to reach Euclidean Rindler space. \\
The cigar background one starts with is of the following form:
\begin{align}
\label{lowest}
ds^2 &= \frac{\alpha'k}{4}\left(dr^2 + 4\tanh^2\left(\frac{r}{2}\right)d\theta^2\right), \\
\label{lowest2}
\Phi &= - \ln\cosh\left(\frac{r}{2}\right).
\end{align}
For bosonic strings, this background receives corrections in $\alpha'$, but for type II superstrings it is $\alpha'$-exact. The string spectrum on this background was determined some time ago, either by looking at the poles of correlation functions \cite{Giveon:1999px}\cite{Giveon:1999tq}\cite{Aharony:2004xn} or by exactly computing the torus path integral \cite{Hanany:2002ev}\cite{Israel:2004ir}. Our strategy in \cite{Mertens:2013zya} was to look at how this string spectrum behaves in the large $k$ limit. Besides looking into the conformal weights of the primaries, we also considered the geometrical interpretation of the spectrum as follows. String fluctuations on this background satisfy the on-shell relation (for type II superstrings):
\begin{equation}
\label{ons}
(L_{0} + \bar{L}_0 - 1) \left|T\right\rangle = 0.
\end{equation}
It is known that one can rewrite the operators $L_0$ and $\bar{L}_0$ in terms of the Laplacian on the coset manifold \cite{Dijkgraaf:1991ba}. The fluctuation hence satisfies the (minimally coupled) Klein-Gordon equation in a curved spacetime in the above metric and dilaton background. \\

\noindent Without going into details, also a dual background was obtained \cite{Dijkgraaf:1991ba} whose geometry determines what winding strings (around the cigar) experience. For type II superstrings, the equation of motion for winding strings is derived from the following action ($L=\sqrt{k\alpha'}$, $\rho = \frac{\sqrt{\alpha' k}}{2}r$)
\begin{align}
S = \int_{0}^{+\infty}d\rho\frac{L}{2}\sinh\left(\frac{2}{L}\rho\right)\left[\left|\partial_\rho T\right|^2 + w^2\frac{\beta^2}{4\pi^2\alpha'^2}\tanh^2\left(\frac{\rho}{L}\right)TT^{*} - \frac{2}{\alpha'}TT^{*} \right],
\end{align}
and this is indeed the naive lowest order in $\alpha'$ action for non-self-interacting fluctuations of winding strings $T$. The large $k$ limit (keeping $\rho$ fixed)gives us then\footnote{A rescaling of $\beta$ has been performed at this step. See \cite{Mertens:2013zya} for details.}
\begin{align}
\label{thscact}
S = \int_{0}^{+\infty}d\rho\rho\left[\left|\partial_\rho T\right|^2 + w^2\frac{\beta^2\rho^2}{4\pi^2\alpha'^3}TT^{*} - \frac{2}{\alpha'}TT^{*} \right].
\end{align}
Taking $k$ large decreases the curvature and flattens the cigar. \\

\noindent The $w=\pm1$ state is special since it determines the dominant near-Hagedorn thermodynamics of the string gas. The reason is that this mode is temperature-dependent and expected to be the least massive of the winding modes. Taking the above (non-interacting) field theory of the thermal scalar and integrating by parts, one can write it schematically as
\begin{equation}
\label{operatorO}
S_{\text{th.sc.}} \sim \int dV e^{-2\Phi}\sqrt{G}T^* \hat{\mathcal{O}}T,
\end{equation}
from which the dominant part of the free energy follows as:
\begin{equation}
\beta F \approx \text{Tr}\text{ln}\hat{\mathcal{O}}.
\end{equation}
For a discrete spectrum of $\hat{\mathcal{O}}$, we obtain
\begin{equation}
\beta F \approx \sum_n\text{ln}\lambda_n
\end{equation}
and it is the lowest eigenmode of $\hat{\mathcal{O}}$ that determines the critical behavior. It was found that in the Euclidean Rindler geometry (\ref{snorm}) (at $\beta = 2\pi\sqrt{\alpha'}$), the lowest eigenmode has the eigenmode and eigenvalue
\begin{equation}
\label{gs}
\psi_0 \propto \exp\left(-\frac{\rho^2}{2\alpha'}\right), \quad \lambda_0=0.
\end{equation}
This means the thermal scalar mode is localized at string length from the horizon and the Rindler temperature (needed to avoid a conical singularity in the geometry) is precisely equal to the Hagedorn temperature $\beta_H = \beta_R$. Since the Hagedorn temperature is associated to the random walking phenomenon of the long string(s) which is identified with the thermal scalar paths, we deduce that the most dominant contribution of the free energy at one loop is given by a random walk at string length from the horizon. \\
For geometries with horizons, in principle the temperature is fixed and one does not have the freedom to change it as innocently as in for instance flat space. String theory has the added difficulty that it is unclear how to deal with general conical spaces. \\
However, the field theory of the thermal scalar does not have this difficulty and one is free to change the temperature. For specific temperatures $\beta = \frac{2\pi\sqrt{\alpha'}}{N}$ (with $N \in \mathbb{N}$), string theory manages to be on-shell and one can make sense of string theory on such spaces. We found that the thermal scalar found on such spaces, agrees with simply taking $\beta = \frac{2\pi\sqrt{\alpha'}}{N}$ in the thermal scalar action (\ref{thscact}) (both for bosonic and for type II superstrings). For heterotic strings, our understanding is more rudimentary although the heterotic analog of (\ref{thscact}) does give the correct dominant mode. We hope to fill this gap in what follows. \\
A further puzzling feature is that modes that have $\left|w\right| > 1$ apparently are not even present in the thermal spectrum. This raises some questions regarding the thermodynamic interpretation of this theory, as usually these are attributed to corrections to Maxwell-Boltzmann statistics \cite{Mertens:2014cia}. \\
Another peculiar part of this story is that (for fully compact spacetimes, as we are instructed to study in thermodynamics), the thermodynamical quantities in principle diverge at the Hagedorn temperature (which equals the Rindler or Hawking temperature). This implies for instance an infinite free energy. One might think that higher genus corrections to the free energy could cure this behavior. In this paper, we will explore this feature more thoroughly. \\
Our goal in this paper is to further utilize the link between the cigar model and Euclidean Rindler space to understand better all of these strange features.

\section{Relevance of long strings for black hole thermodynamics}
\label{relevance}
First let us ask a general question: can one ignore the massive string modes when computing loop corrections to thermodynamical quantities in black hole spacetimes? Hence we wish to contemplate whether approximating the one loop string free energy by only the free energy of the massless modes (in the Lorentzian spectrum) (i.e. gravitons, photons etc.) is a good approximation. \\
String loop corrections to black hole thermodynamics have a long history riddled with controversy, see e.g. \cite{Susskind:1994sm}\cite{Susskind:1993ws}\cite{Dabholkar:1994ai}\cite{Lowe:1994ah}\cite{Susskind:2005js}\cite{Parentani:1989gq}\cite{Emparan:1994bt}\cite{McGuigan:1994tg}. In general one expects higher worldsheet corrections to be neglible when considering the exterior of black holes. The argument is well-known: higher worldsheet corrections (or massive string modes) manifest themselves in the low energy effective action as corrections of higher order in $\alpha'/R^2$ with $R$ some curvature radius. The curvature outside a (large) black hole horizon is much smaller than the inverse string length. Hence these corrections are very small and can be neglected, suggesting the massless modes give the main contribution to observables for large black holes. \\

\noindent The situation is completely different however when considering one-loop thermodynamical quantities. This can be appreciated from different perspectives.\\
A first primitive argument is as follows. We noted previously in \cite{Mertens:2013zya} that for Rindler space plus a fully compact remainder, the free energy itself diverges as $\beta F = \ln(\beta-\beta_H)$ where $\beta_H = \beta_{\text{Hawking}}$ and one should set $\beta$ equal to the Hawking temperature as well in the end. This implies $F$ diverges on the nose. This is obviously not achieved by only considering the massless fields around the black hole.\footnote{A divergence sets in in this case as well, though it is temperature-independent.} \\
For a different argument, consider the thermal partition function. The radius of the thermal circle is an extra curvature parameter. We have explicitly demonstrated elsewhere \cite{Mertens:2013zya} that higher order $\alpha'$ corrections constructed with the inverse temperature $\beta$ are \emph{not} subdominant for thermal winding modes. \\

\noindent A more physical point of view can be given on the Lorentzian signature manifold.\footnote{Although we are somewhat reluctant to have too much faith in it due to the comments in the next footnote.} To that effect, let us first look at the formulas for the flat space string. The free energy of a (bosonic) field of mass $m$ in $D+1$ dimensions is given by
\begin{equation}
\beta F = V\int\frac{d^{D}k}{(2\pi)^D} \ln\left(1-e^{-\beta\sqrt{k^2+m^2}}\right).
\end{equation}
Clearly a higher mass field has a lower free energy. In the large mass limit, we can approximate
\begin{equation}
\ln(1-x) \approx -x,
\end{equation}
which makes the integrand proportional to $\propto e^{-\beta E}$. When considering string theory, we should multiply this by the degeneracy of states $\propto e^{\beta_H E}$ (for large mass). We conclude that for $T \ll T_H$, the lowest $m^2$ modes give the largest contribution to the free energy: the degeneracy of high $m^2$ states cannot compete with the lower mass. For $T \lesssim T_H$, the higher $m^2$ modes are not subdominant but give important contributions to the free energy: the full string theory is relevant. \\
For black holes, the above computation goes through almost identically.\footnote{Note though that this has been questioned in \cite{Susskind:1994sm} in the following way. It was suggested that the genus one result on the thermal manifold does \emph{not} correspond to the free-field trace, but instead includes some interactions with open strings whose endpoints are fixed on the horizon. Despite being an explicit proposal on the stringy microscopic degrees of freedom, little success has been booked in using and/or proving aspects of this proposal since then. We will nonetheless assume that the free-field Hamiltonian trace has the same critical Hagedorn temperature as predicted by genus one string thermodynamics. The reason is that we believe we have given evidence that the critical temperature of the thermal scalar is tightly linked to the random walk phenomenon close to black hole horizons and this long string picture is precisely what is expected near the horizon on general arguments \cite{Susskind:1993ws}\cite{Susskind:1993ki}\cite{Susskind:1994uu}\cite{Susskind:1993aa}, strongly suggesting the equality of $\beta_H$ and $\beta_{\text{Hawking}}$ also for the free-field trace. Note that this assumption is only made in this argument here and all other results that follow in other sections are independent of its validity. We will in fact further investigate this issue in section \ref{modDom}.} The free energy of a non-interacting Bose (and/or Fermi) gas of strings is given by
\begin{equation}
\beta F =  \pm \sum_{\text{species}}\ooalign{$\displaystyle\sum$\cr\hidewidth$\displaystyle\int$\hidewidth\cr}_{E_i} \ln\left(1 \mp e^{-\beta E_i}\right),
\end{equation}
where we sum over all Lorentzian string states in the spectrum. The high energy states again provide a factor of $e^{-\beta E}$, with $\beta$ the inverse Hawking temperature. Since this precisely coincides with the Hagedorn temperature (determining the degeneracy of high energy string states), the highly excited modes are very relevant and it is incorrect to approximate the one-loop free energy of strings by that given solely by the massless modes. \\
In the physical picture we have, the massless modes alone do not give the random walker surrounding the horizon; this is only obtained by considering the highly excited strings. This is to be contrasted with several holographic computations (e.g. \cite{Denef:2009yy} where the authors compute the one-loop free energy in a holographic black hole background using only one class of charged matter).\footnote{Of course we do not claim in any way that these authors are wrong, we simply point out some tension between the gravity-plus-matter approach and the full string picture at one loop. Moreover, these authors consider charged black holes while in our case uncharged black holes are studied.}

\section{Approach to Heterotic Euclidean Rindler space}
\label{heterotic}
In this section we look at the near-horizon Rindler approximation of black holes. In \cite{Mertens:2013zya} we analyzed the critical one-loop string thermodynamics and found the following results for the non-interacting thermal scalar field theory. The thermal scalar action for type II superstrings is exactly given by the lowest order (in $\alpha'$) action as was shown in \cite{Giveon:2012kp}\cite{Giveon:2013ica}\cite{Giveon:2014hfa} by taking the large $k$ limit of the $SL(2,\mathbb{R})/U(1)$ black hole. The bosonic string thermal scalar on the other hand does contain $\alpha'$ corrections. We also observed that heterotic string theory on flat $\mathbb{C}/\mathbb{Z}_N$ orbifolds agrees with a thermal scalar action without any $\alpha'$ corrections (like for type II superstrings), but we were unable to give a proof of this statement. In this section we will present an argument as to why this is so. As in \cite{Giveon:2013ica}\cite{Mertens:2013zya}, we are looking for a suitable cigar CFT to take the large $k$ limit. There exist several approaches and points of view on heterotic coset models (see e.g. \cite{Johnson:1994jw}\cite{Johnson:1994kv}\cite{Johnson:2004zq}\cite{Svendsen:2005gy}\cite{Giveon:1993hm}\cite{Sfetsos:1993bh}), and also several realizations of the analog of the cigar CFT. We choose the left-right symmetric realization where the heterotic worldsheet theory actually has (1,1) supersymmetry instead of the expected (0,1) supersymmetry \cite{Giveon:1993hm}. Heterotic backgrounds can be trivially constructed from type II backgrounds by embedding the spin connection in the gauge connection. This approach was explored by \cite{Giveon:1993hm} to discuss heterotic WZW models. One of the benefits of this approach is that the techniques from type II coset models can be integrally carried over to this case, in particular the identification of the exact background fields and the resulting ($\alpha'$-exact) tachyon equation of motion. \\
For more general heterotic models it becomes less clear whether such an approach is viable. Other methods to distill the metric and dilaton exist in this case \cite{Johnson:2004zq}, but we also need to determine the tachyon equation of motion, and the approach followed in \cite{Dijkgraaf:1991ba} is ideally suited for this. \\

\noindent As a motivation to consider the left-right symmetric models, we note the following. In general, the Busher rules for heterotic strings receive $\alpha'$-corrections. Hence the thermal scalar action in heterotic string theory receives $\alpha'$-corrections, just like the bosonic string. However, if the background has an enlarged supersymmetry compared to the expected $(0,1)$ SUSY, the heterotic string effectively behaves as a type II superstring and the Busher rules do not get corrections (at least for (gauged) WZW models). The fact that for Rindler space the lowest order (in $\alpha'$) thermal scalar action appears to be $\alpha'$-exact, is evidence that in this case indeed more supersymmetry is present than expected. This shows why the left-right symmetric approach to heterotic coset models (effectively giving type II models), is the most natural place to look for realizing Euclidean Rindler space in heterotic string theory. \\

\noindent Thus to any type II background, one can associate a heterotic background by embedding the spin connection into the gauge connection. For this heterotic background, the fluctuation equations of the states is given by the same $L_0$ and $\bar{L}_0$ (written in terms of the Laplacian on the coset) as for the type II superstring \cite{Giveon:1993hm}. The only difference is in the precise on-shell conditions, which for heterotic strings are given by
\begin{align}
L_0 - 1 &=0 ,\\
\bar{L}_0 - 1/2 &= 0.
\end{align}

\begin{figure}[h]
\begin{minipage}{0.5\textwidth}
\centering
\includegraphics[width=5cm]{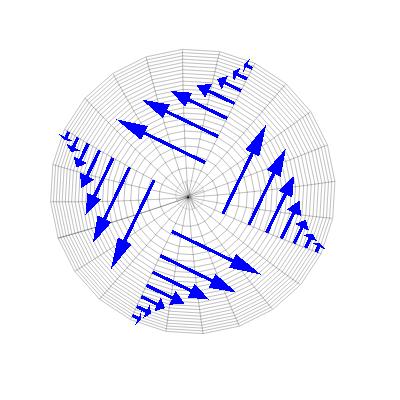}
\end{minipage}
\begin{minipage}{0.5\textwidth}
\centering
\includegraphics[width=5cm]{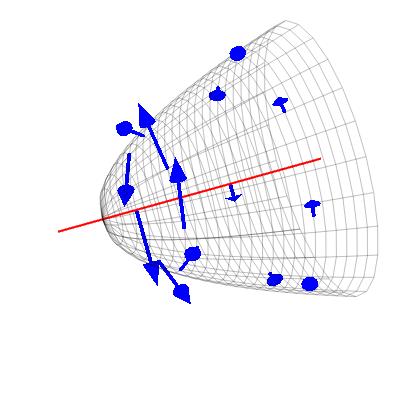}
\end{minipage}
\caption{Left figure: top view of the cigar with the background gauge field schematically shown. Right figure: side view of the cigar and the background gauge field.}
\label{topside}
\end{figure}

\noindent Within such a left-right symmetric approach, a $SL(2,\mathbb{R})/U(1)$ CFT can be found with the following background fields \cite{Giveon:1993hm}
\begin{align}
\label{bgfields1}
ds^2 &= \frac{\alpha'k}{4}\left(dr^2 + 4 \tanh\left(\frac{r}{2}\right)^2 d\theta^2\right), \\
\label{bgfields2}
\Phi &= \Phi_0 - \ln\left(\cosh\left(\frac{r}{2}\right)\right), \\
\label{bgfields3}
A_\theta &= -\frac{1}{\cosh\left(\frac{r}{2}\right)^2},
\end{align}
where the gauge connection equals the Lorentz spin connection and $A_r = 0$. The spin connection is valued in the holonomy group of the 2d space (being $U(1)$). Hence the result is an Abelian gauge field $A_\mu$ that resides in some $U(1)$ subalgebra of the full heterotic gauge algebra. The angular coordinate is identified as $\theta \sim \theta + 2\pi$. These coordinates are however singular for $r=0$, in the same way that polar coordinates are. Normally one can readily continue the solutions to include $r=0$, since no physical singularities are encountered at $r=0$. In this case however, the gauge field becomes singular at $r=0$, and this singularity has physical consequences. We therefore conclude that the above solution is only valid for $r>0$. The gauge field and its singular character are schematically depicted in figure \ref{topside}.

\noindent The dual background (corresponding to a $U(1)$ vector gauging) is given by
\begin{align}
\label{dualbg}
ds^2 &= \frac{\alpha'k}{4}\left(dr^2 + 4 \coth\left(\frac{r}{2}\right)^2 d\tilde{\theta}^2\right), \\
\tilde{\Phi} &= \tilde{\Phi_0} - \ln\left(\sinh\left(\frac{r}{2}\right)\right), \\
\tilde{A}_{\tilde{\theta}} &= \frac{1}{\sinh\left(\frac{r}{2}\right)^2},
\end{align}
where now $\tilde{\theta} \sim \tilde{\theta} + \frac{2\pi}{k}$. These background fields determine how winding modes sense the geometry. Like for the bosonic and type II case, this geometry is trumpet-shaped with a curvature singularity at the origin $r=0$. This background is $\alpha'$-exact and the tachyon equation of motion can be readily determined in this background. Like for the type II string, the winding tachyon equation of motion does not get $\alpha'$ corrections compared to the T-dual geometry of the background.\footnote{The T-dual metric, dilaton and Kalb-Ramond field are not influenced by the non-vanishing gauge field.} In this case, the dual gauge field also blows up at the origin, but this is irrelevant for the thermal scalar equation of motion since this one is only determined by the dual metric and dual dilaton. Writing down the discrete momentum tachyon equation in this background (\ref{dualbg}) gives us the winding tachyon equation of motion we are after. \\
Note that the propagation equations for the thermal scalar in this background are constructed from the same ingredients as those of the type II superstring (the $L_0$ and $\bar{L}_0$ operators have the same form) and hence the background gauge field does not couple directly to the tree-level action of the thermal scalar. This in particular seems to imply that the thermal scalar does not carry any charge corresponding to this background gauge field. \\
To see the flat limit, we substitute $\rho = \frac{\sqrt{\alpha' k}}{2} r$ and then take $k \to \infty$ keeping $\rho$ fixed. The geometry reduces to polar coordinates, the dilaton becomes constant and the gauge field also becomes constant.\\

\noindent To proceed, we first discuss the singular character of the background fields. Let us briefly look at the nature of the gauge field near the horizon. Near $r=0$, the geometry reduces to polar coordinates, and the gauge field becomes constant. So we are actually interested in a constant angular gauge field $A_\theta = C = -1$ in polar coordinates. One readily computes the Cartesian components as
\begin{equation}
\label{CS}
A_x = C\frac{d\theta}{dx}= -\frac{Cy}{x^2+y^2}, \quad A_y = C \frac{d\theta}{dy} = \frac{Cx}{x^2+y^2}
\end{equation}
and one finds that the field tensor $F_{xy}$ vanishes (almost) everywhere (or directly in polar coordinates $F_{\rho\theta} =0$).
Nevertheless, the flux does not vanish since
\begin{equation}
\int_{\text{disc}} F = \oint_{\text{circle}}d\theta A_{\theta} = 2\pi C
\end{equation}
so we find $F_{xy} = 2\pi C \delta(x,y)$ and there is a delta-source of magnetic flux present at the origin. This flux is important because charged states can acquire an Aharonov-Bohm phase upon circling the origin. From another perspective, the singularity of the gauge field at the origin is translated to the violation of the commutativity of partial derivatives of the angular coordinate at the origin:
\begin{equation}
F_{xy} = C\left(\partial_x\partial_y - \partial_y\partial_x \right)\theta \neq 0 \quad \text{at }r=0.
\end{equation}

\noindent Our strategy is to perform a (large) gauge transformation to eliminate the gauge field. The gauge transformation has two effects that need to be separately analyzed. Firstly, the other background fields might change, undoing the very thing we try to accomplish. This can be analyzed easiest by turning to the effective spacetime action: a solution of the all-order in $\alpha'$ effective field theory yields a consistent background for string propagation.\\
Secondly, the fluctuations on this background might be charged under the background gauge field. They hence can feel this gauge transformation. Both of these will be looked at now.

\subsection{Effect of gauge transformation on background}
The background gauge transformation can influence the background fields. To lowest order in $\alpha'$, it is known that the Kalb-Ramond 2-form (despite being uncharged under $A_\mu$) undergoes a simultaneous gauge transformation of the form:
\begin{align}
\delta B_2 &\propto \text{Tr}\left(\lambda dA\right), \\
\delta A &= d\lambda.
\end{align}
More generally, at higher orders in $\alpha'$, the $B_2$-form is known to transform also under gauge transformations of the Lorentz connection, though we will not need this. Note that the only (massless) field that \emph{is} charged under the background gauge field is the gaugino. This field transforms under a gauge transformation in the adjoint representation of the gauge group (homogeneously) and hence if it is turned off initially, it will not be turned on by a gauge transformation. \\
What is crucial is that this gauge transformation is dictated by the Green-Schwarz anomaly cancellation mechanism \cite{Green:1984sg}, and does \emph{not} depend on the concrete form of the spacetime effective action. Hence the above gauge transformation should hold to all orders in $\alpha'$. And indeed, the analysis of \cite{Callan:1985ia}\cite{Metsaev:1986yb}\cite{Foakes:1987bn} shows that, up to three loops on the worldsheet, it is only the gauge-independent combination 
\begin{equation}
\tilde{H}_3 = dB_2 -c \omega_{3Y} - c'\omega_{3L}
\end{equation}
that appears in the effective action. In this formula, the $\omega_3$'s are the Chern-Simons 3-forms constructed with the Yang-Mills connection $A$ or the Lorentz connection. \\
For Rindler space, we already have that $dA = 0$ for $r\neq 0$. This means $B_2$ does not become non-zero for $r\neq0$ after the gauge transformation. Note that this is very different from the original $A$. The original $A$ field had a constant angular component and hence a singularity at the origin. The new $B_2$ is zero everywhere and can be chosen zero at the origin as well: no global analysis (such as a line integral for $A$) can detect something is present at $r=0$. For all intents and purposes, the $B_2$-form is absent. This gauge transformation obviously maps solutions to solutions, and hence we can safely turn off the background $A$-field in this case: no effects on other background fields are present.

\subsection{Effect of gauge transformation on fluctuations}
\begin{figure}[h]
\centering
\includegraphics[width=5cm]{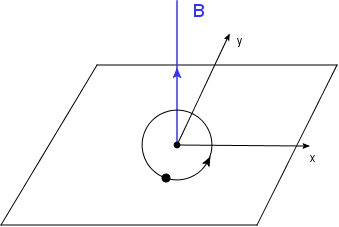}
\caption{The $(\rho, \theta)$ plane (or the $xy$ plane) and the delta-source of magnetic flux at the origin. States that circle the origin can receive an arbitrary phase factor corresponding to anyon statistics.}
\label{cart}
\end{figure}

Gauge fields of the form (\ref{CS}) are well-known from studies of matter-coupled Chern-Simons gauge theories in (2+1) dimensions and the related anyon statistics (see e.g. \cite{Dunne:1998qy} and references therein). In the $k\to\infty$ limit, the gauge field is pure gauge for $r \neq 0$ and can be eliminated. However, charged states get multiplied by an angle-dependent prefactor and their periodicity or anti-periodicity upon circling the origin gets altered to general anyon statistics. A sketch of the situation is given in figure \ref{cart}.
In somewhat more detail, we can write (for $i=1,2$, the Cartesian coordinates in the ($\rho,\theta$) plane)
\begin{equation}
A_i = C \partial_i \arg(\mathbf{x}) = -\partial_i \theta.
\end{equation}
Under a gauge transformation with $A_i \to A_i + \partial_i \chi$ where $\chi = - C \arg(\mathbf{x})$, the wavefunction of a charged state gets changed into
\begin{equation}
\psi \to e^{ie\chi} \psi = e^{ie\theta} \psi,
\end{equation}
which hence changes indeed the phase of the state upon circling the origin by $e^{2\pi i e}$. The charges of the matter depend on which $U(1)$ embedding is chosen and it is difficult to say anything concrete about these at this point.\footnote{Life would be simpler if we could choose an Abelian ideal instead, but unfortunately a semi-simple algebra does not have such ideals.} \\
Let us now return to the entire cigar. We perform a large gauge transformation that kills the singularity of the gauge field at the horizon. This introduces a non-zero Wilson loop at infinity. This kind of reasoning was performed also in \cite{Giveon:2004rw} in a different context.\footnote{See also \cite{Giveon:2005jv} for more discussions on this.} Charged string states hence `feel' the gauge transformation. In our case, we are interested in the thermal scalar. This stringy state is uncharged under the gauge field\footnote{This is because the gauge field comes from the 10 dimensional heterotic gauge field. If for example the gauge field is actually a component of the Kaluza-Klein reduced metric, the thermal scalar would be charged and the above reasoning would not hold. In fact, the thermal scalar of the heterotic string is explicitly charged under the Kaluza-Klein gauge field $G_{\mu \tau}$ and under the Kalb-Ramond gauge field $B_{\mu\tau}$, as one can for instance see very explicitly in the low energy field theory of the thermal scalar as in \cite{Schulgin:2011zb}.} (it only couples to the gauge field through $F_{\mu\nu}$ at higher orders in $\alpha'$) and hence the statistics is unchanged (i.e. it does not get premultiplied by a Wilson loop upon circling the cigar). So after performing this (large) gauge transformation, the resulting equation of motion of the thermal scalar is the same as before. \\

\noindent We summarize what we have done so far: the above background (equations (\ref{bgfields1})-(\ref{bgfields3})) is valid only for $r \neq 0$. For this background the exact fluctuation equations for the string states are known. We then perform a large gauge transformation that eliminates the gauge singularity at $r=0$. This transformation has physical consequences by introducing a non-trivial Wilson-loop that influences charged string states. The thermal scalar on the other hand is uncharged and its equation of motion is not influenced by the large gauge transformation. Finally taking $k\to\infty$ while keeping $\rho = \frac{\sqrt{k\alpha'}r}{2}$ fixed, we recover Euclidean Rindler space for which the fluctuation equations for uncharged states have remained the same during the gauge transformation. The fate of the charged string states is another question, but for the purposes of this paper we are only interested in the thermal scalar itself. For uncharged states, the gauge field has no physical impact anymore and we effectively reduce the model to Euclidean Rindler space. \\

\noindent The propagation equations for the heterotic string are
\begin{align}
L_0 - 1 &=0 ,\\
\bar{L}_0 - 1/2 &= 0.
\end{align}
Adding and subtracting gives
\begin{align}
L_0 +\bar{L}_0- 3/2 &=0 ,\\
L_0 - \bar{L}_0 - 1/2 &= 0.
\end{align}
For this left-right symmetric cigar CFT the conformal weights reduce to those of the type II superstring and so we find the conformal weights:\footnote{The spectrum on the cigar CFT is explicitly written down in appendix \ref{cigspectr} for bosonic and type II superstrings. The reader who is not aware of these formulas should consult it at this point.}
\begin{equation}
h = -\frac{j(j-1)}{k} + \frac{m^2}{k}, \quad \bar{h} =  -\frac{j(j-1)}{k} + \frac{\bar{m}^2}{k},
\end{equation}
so the physicality constraint becomes
\begin{equation}
h - \bar{h} = wn = 1/2,
\end{equation}
just like in flat space. The equation of motion is determined by writing $L_0 + \bar{L}_0$ in terms of the Casimir operator, and is exactly the same as for type II superstrings. The metric one obtains is again:
\begin{equation}
ds^2 = \frac{\alpha' k}{4}\left[dr^2 + \frac{4}{\coth^2\left(\frac{r}{2}\right)}d\theta^2 + \frac{4}{\tanh^2\left(\frac{r}{2}\right)}d\tilde{\theta}^2\right].
\end{equation}
The eigenvalue equation for the NS primaries, that one obtains by taking $k\to\infty$, is now:\footnote{At least for the uncharged string states as discussed above.}
\begin{equation}
-\frac{\partial_\rho\left(\rho\partial_{\rho}T(\rho)\right)}{\rho}+ \left[-\frac{3}{\alpha'} + n^2\frac{1}{\rho^2} + w^2\frac{\rho^2}{\alpha'^2} \right]T(\rho)= \lambda T(\rho).
\end{equation}
The thermal scalar equation hence becomes
\begin{equation}
\left[-\partial_\rho^2  - \frac{1}{\rho}\partial_{\rho} - \frac{3}{\alpha'} + \frac{1}{4\rho^2} + \frac{\rho^2}{\alpha'^2} \right]T(\rho)= \lambda T(\rho).
\end{equation}
From this we conclude that indeed heterotic strings also do not receive $\alpha'$ corrections to the (quadratic part of the) thermal scalar action. The physicality constraint is the same as in flat space, and the thermal scalar action combines discrete momentum and winding around the Rindler origin. We already noted \cite{Mertens:2013zya} that the critical behavior as dictated by this action (and no corrections to it) agrees with the flat space $\mathbb{C}/\mathbb{Z}_N$ orbifold thermodynamics \cite{Dabholkar:1994ai}\cite{Lowe:1994ah}. \\
When considering the entire spectrum, we note that the only difference (modulo the charged states issue) with type II strings is that the constraint is different and so different states are allowed or forbidden. The constraint is
\begin{equation}
N_L - N_{R} + nw = 1/2,
\end{equation}
even before taking the $k\to\infty$ limit. In particular, the wavefunctions of the states are the same as those for type II superstrings. One can imagine that for asymmetric constructions this constraint might be different.

\section{Flat space limit of the cigar partition function}
\label{flatlimit}
In \cite{Giveon:2014hfa} it was shown for the type II superstring that the large $k$ limit can be directly taken at the partition function level. The authors use a modular invariant regularization to deal with the internal divergences and reinterpret this regulator (together with the level $k$) as the volume divergence of flat space. The discrete states are however not found anymore in the large $k$ limit of the partition function and it was speculated that their imprint should be found as a $1/k$ effect. \\
In this section we study this limit further. We first focus on the bosonic string and its orbifolds. The benefit of studying orbifolds is that in the twisted sectors no volume divergence arises and this allows a clean comparison between the flat $\mathbb{C}/\mathbb{Z}_N$ partition function and the cigar orbifold partition function in the large $k$ limit. This allows us to make the link between these models more precise. We also present similar formulas for type II superstrings. For these, the authors of \cite{Giveon:2014hfa} argued that the tip of the cigar is special due to the GSO projection imposed at infinity, essentially causing a breakdown of the coordinate equivalence between cartesian and polar coordinates in the flat limit. We will come back to this point in what follows. In \cite{Mertens:2013zya} we only compared the dominant thermal scalar behavior of both partition functions. The results of this section can be viewed as a more elaborate comparison of the full partition functions. 

\subsection{Bosonic cigar CFT}
The partition function for the $\mathbb{Z}_N$-orbifolded cigar CFT can be written in the following form \cite{Sugawara:2012ag}\cite{Mertens:2013zya}\cite{Hanany:2002ev}:
\begin{align}
\label{starting}
Z &= \frac{1}{N}2\sqrt{k(k-2)}\int_{\mathcal{F}}\frac{d\tau d\bar{\tau}}{\tau_2} \int_{-\infty}^{+\infty}ds_1ds_2 \nonumber \\
&\sum_{m,w=0}^{N-1}\sum_i q^{h_i}\bar{q}^{\bar{h}_i}e^{4\pi\tau_2(1-\frac{1}{4(k-2)}) -\frac{k\pi}{\tau_2}\left|(s_1 - \frac{w}{N})\tau +(s_2 - \frac{m}{N})\right|^2+2\pi\tau_2s_1^2} \nonumber \\
&\frac{1}{\left|\sin(\pi(s_1\tau + s_2))\right|^2}\left|\prod_{r=1}^{+\infty}\frac{(1-e^{2\pi i r \tau})^2}{(1-e^{2\pi i r \tau - 2\pi i (s_1\tau +s_2)})(1-e^{2\pi i r \tau + 2\pi i (s_1\tau +s_2)})}\right|^2.
\end{align}
In this formula, the $bc$-ghosts have already been included and an internal CFT with weights $h_i$ is left arbitrary. As usual, $q=\exp(2\pi i \tau)$. First we focus on the twisted sectors (with $(w, m) \neq (0,0)$), since these do not exhibit a volume divergence. The large $k$ limit implies that the $s_1$- and $s_2$-integrals are dominated by $s_1 = w/N$ and $s_2 = m/N$. The infinite product factor in the end should simply be evaluated at this point. The theta-function appearing here can be directly related to the theta function with characteristics using the following set of formulas (and setting $\nu = \frac{w}{N} \tau + \frac{m}{N}$):
\begin{align}
\theta_1(\nu,\tau) &= 2 e^{\pi i \tau/4}\sin(\pi\nu) \prod_{n=1}^{+\infty}(1-q^n)(1-zq^n)(1-z^{-1}q^n), \\
-\theta_1(\nu,\tau) &= \theta\left[
\begin{array}{c}
\frac{1}{2}\\
\frac{1}{2}\end{array} 
\right](\nu, \tau) = e^{\frac{\pi i \tau}{4} + \pi i (\nu + \frac{1}{2})}\theta\left(\nu+\frac{\tau}{2}+\frac{1}{2},\tau\right), \\
\theta\left[
\begin{array}{c}
\frac{1}{2}+\frac{w}{N} \\
\frac{1}{2}+\frac{m}{N}  \end{array} 
\right](\tau) &= e^{\pi i \tau \left(\frac{1}{2} +\frac{w}{N}\right)^2 + 2\pi i \left(\frac{1}{2}+\frac{w}{N}\right)\left(\frac{1}{2}+\frac{m}{N}\right)} \theta\left(\left(\frac{1}{2}+\frac{w}{N}\right)\tau + \left(\frac{1}{2} + \frac{m}{N}\right), \tau\right),
\end{align}
where $z=\exp(2\pi i \nu)$. After some straightforward arithmetic, we can then write for this sector:
\begin{align}
Z_{w,m} &\approx \frac{1}{N}2\sqrt{k(k-2)}\int_{F}\frac{d\tau d\bar{\tau}}{\tau_2} \int_{-\infty}^{+\infty}ds_1ds_2 \nonumber \\
&\sum_i q^{h_i}\bar{q}^{\bar{h}_i}e^{4\pi\tau_2(1-\frac{1}{4(k-2)}) -\frac{k\pi}{\tau_2}\left|(s_1 - \frac{w}{N})\tau +(s_2 - \frac{m}{N})\right|^2}
\frac{4\left|\prod_{n=1}^{+\infty}(1-q^n)\right|^6 e^{-\pi\tau_2/2}}{\left|\theta\left[
\begin{array}{c}
\frac{1}{2}+\frac{w}{N} \\
\frac{1}{2}+\frac{m}{N}  \end{array} 
\right](\tau)\right|^2}.
\end{align}
To evaluate the integral, we make use of polar coordinates in the form $s_1 \tau + s_2 = x_1 + ix_2$ or\footnote{To be more precise, we first shift $s_1 \to s_1 + w/N$ and $s_2 \to s_2 + m/N$.} 
\begin{align}
s_1 \tau_1 + s_2 &= \rho \cos(\phi), \\
s_1 \tau_2 &= \rho \sin(\phi).
\end{align}
The transformation from ($s_1$, $s_2$) to ($\rho$, $\phi$) has Jacobian $\rho/\tau_2$. Hence the remaining integral becomes
\begin{equation}
\frac{2\pi}{\tau_2} \int_{0}^{+\infty}d\rho \rho e^{-\frac{\pi k}{\tau_2}\rho^2} = \frac{1}{k}.
\end{equation}
In the large $k$-limit, we hence obtain
\begin{align}
Z_{w,m} \approx \frac{1}{N}2\int_{F}\frac{d\tau d\bar{\tau}}{\tau_2} 
\sum_i q^{h_i}\bar{q}^{\bar{h}_i}e^{4\pi\tau_2}\frac{4\left|\eta(\tau)\right|^6}{\left|\theta\left[
\begin{array}{c}
\frac{1}{2}+\frac{w}{N} \\
\frac{1}{2}+\frac{m}{N}  \end{array} 
\right](\tau)\right|^2},
\end{align}
where $\eta(\tau) = q^{1/24}\prod_{n=1}^{+\infty}(1-q^n)$ is the Dedekind eta-function. Note that the factor $e^{4\pi\tau_2} = (q\bar{q})^{-1}$ is to be interpreted as the central charge term of the internal CFT with $c=24$. Choosing this internal CFT to be flat, we obtain in the end
\begin{align}
Z_{w,m} &\approx \frac{V_T}{N}2\int_{F}\frac{d\tau d\bar{\tau}}{\tau_2}
\frac{1}{(4\pi^2\alpha'\tau_2)^{12}}\left|\eta\right|^{-48}\frac{4\left|\eta\right|^6}{\left|\theta\left[
\begin{array}{c}
\frac{1}{2}+\frac{w}{N} \\
\frac{1}{2}+\frac{m}{N}  \end{array} 
\right](\tau)\right|^2}.
\end{align}
which agrees with the flat ($w$, $m$) sector \cite{Dabholkar:1994ai}\cite{Lowe:1994ah}.\footnote{In fact, we are off by a factor of 32. We would like to have obtained instead
\begin{align}
Z_{w,m} &\approx \frac{V_T}{N}\int_{F}\frac{d\tau d\bar{\tau}}{4\tau_2}
\frac{1}{(4\pi^2\alpha'\tau_2)^{12}}\left|\eta\right|^{-48}\frac{\left|\eta\right|^6}{\left|\theta\left[
\begin{array}{c}
\frac{1}{2}+\frac{w}{N} \\
\frac{1}{2}+\frac{m}{N}  \end{array} 
\right](\tau)\right|^2}.
\end{align}
We interpret this as a factor that should be included in the result of \cite{Hanany:2002ev}. For the type II superstring, the expressions given in the literature have a different normalization, and we will not have this discrepancy anymore. In the following computation of the bosonic string, we include this factor of $1/32$.} \\ 

\noindent The sector $w=m=0$ should be dealt with separately, since the sine function in (\ref{starting}) causes a divergence that is to be interpreted as a IR volume divergence. \\
The above analysis has the following modifications. Firstly, the saddle point is at $s_1=s_2=0$. This implies the infinite product in (\ref{starting}) becomes equal to 1. The sine factor blows up at the origin and we regulate it by cutting out a $\epsilon$-sized circle in the $x_1, x_2$ plane (following \cite{Giveon:2014hfa}). The final saddle-point integral is given by\footnote{In writing this we used the formula for the exponential integral: 
\begin{equation}
\int dx \frac{e^{-Ax^2}}{x} = -\frac{1}{2} \text{Ei}(1,Ax^2),
\end{equation}
with the series expansion
\begin{equation}
\label{taylor}
\text{Ei}(1,Ax^2) \approx -\gamma -\ln(A) -2\ln(x) + Ax^2 - \frac{1}{4} A^2 x^4 + \mathcal{O}(x^6).
\end{equation}
The first two terms are irrelevant in a definite integral (such as the one we have here). The third term is important: it gives precisely the $\ln(\epsilon)$ dominant contribution.}
\begin{equation}
\frac{2\pi}{\tau_2}\int_{\epsilon}^{+\infty}d\rho \rho \frac{e^{-\frac{\pi k }{\tau_2}\rho^2}}{\pi^2\rho^2} = -\frac{2}{\pi \tau_2} \ln(\epsilon).
\end{equation}
The resulting expression can be compared with the $w=m=0$ sector of $\mathbb{C}/\mathbb{Z}_N$ which is given by
\begin{equation}
\label{flatbosonic}
Z = \frac{V_T \mathcal{A}}{N}\int_{F}\frac{d\tau d\bar{\tau}}{4\tau_2^2}\frac{1}{(4\pi^2\alpha'\tau_2)^{12}}\left|\eta(\tau)\right|^{-48}.
\end{equation}
This allows us to identify the transverse area with the following divergent quantities:
\begin{equation}
\mathcal{A} = -\frac{1}{2\pi}\sqrt{k(k-2)}\ln(\epsilon).
\end{equation}
In the large $k$ limit this becomes\footnote{This implies that for fixed area $\mathcal{A}$, $k$ scales as $-\frac{1}{\ln(\epsilon)}$. The higher terms in the above expansion (\ref{taylor}) are then of the form (including the $k$ prefactor present in the partition function (\ref{starting}) itself)
\begin{equation}
k^2 \epsilon^2 \sim  \frac{\epsilon^2}{\ln(\epsilon)^2} \to 0,
\end{equation}
or for the general term
\begin{equation}
\frac{\epsilon^{2n}}{\ln(\epsilon)^{n+1}} \to 0.
\end{equation}
Hence there are no subleading corrections that survive the $k\to\infty$ limit.}
\begin{equation}
\label{area}
\boxed{
\mathcal{A} = -\frac{1}{2\pi}k\ln(\epsilon)}.
\end{equation}

\subsection{Extension to the type II superstring for odd $N$}
The above reasoning can be readily generalized to the $\mathbb{Z}_N$ orbifolds of the type II superstring on each of these spaces. The formulas are a bit long, but the logic is the same as for the bosonic string. For odd $N$, the flat conical partition function is of the form \cite{Lowe:1994ah}
\begin{align}
\label{flatpt}
Z(\tau) = \frac{1}{4N}\left(\frac{1}{\left|\eta\right|^2\sqrt{4\pi^2\alpha'\tau_2}}\right)^{6}\sum_{w,m=0}^{N-1}&\sum_{\alpha,\beta,\gamma,\delta}\omega'_{\alpha\beta}(w,m)\bar{\omega}'_{\gamma\delta}(w,m) \nonumber \\
&\times \frac{\theta\left[
\begin{array}{c}
\alpha \\
\beta  \end{array} 
\right]^3 \theta\left[
\begin{array}{c}
\alpha+\frac{w}{N} \\
\beta+\frac{m}{N}  \end{array} 
\right]\bar{\theta}\left[
\begin{array}{c}
\gamma \\
\delta  \end{array} 
\right]^3 \bar{\theta}\left[
\begin{array}{c}
\gamma+\frac{w}{N} \\
\delta+\frac{m}{N}  \end{array} 
\right]}{\left|\theta\left[
\begin{array}{c}
\frac{1}{2}+\frac{w}{N} \\
\frac{1}{2}+\frac{m}{N}  \end{array} 
\right]\eta^3\right|^2}.
\end{align}
The $\omega'$ prefactors are given as follows
\begin{align}
\omega'_{00}(w,m) &= 1 , \\
\omega'_{0\frac{1}{2}}(w,m) &= e^{-\frac{\pi i w}{N}}(-1)^{w+1}, \\
\omega'_{\frac{1}{2}0}(w,m) &= (-1)^{m+1}, \\
\omega'_{\frac{1}{2}\frac{1}{2}}(w,m) &= \pm e^{-\frac{\pi i w}{N}}(-1)^{w+m}.
\end{align}
The partition function on the cigar orbifold on the other hand is given by \cite{Sugawara:2012ag}\cite{Giveon:2014hfa}
\begin{align}
\label{cigarpt}
Z(\tau) = &\frac{k}{N} \sum_{\sigma_L, \sigma_R}\sum_{w,m \in \mathbb{Z}} \int_{0}^{1}ds_1ds_2 \epsilon(\sigma_L;w,m)\epsilon(\sigma_R;w,m) \nonumber \\
&\times f_{\sigma_L}(s_1\tau+s_2,\tau)f_{\sigma_R}^{*}(s_1\tau+s_2,\tau)e^{-\frac{\pi k}{\tau_2}\left|\left(s_1 - \frac{w}{N}\right)\tau + \left(s_2 - \frac{m}{N}\right)\right|^2}.
\end{align}
where 
\begin{equation}
f_{\sigma}(u,\tau) = \frac{\theta_{\sigma}(u,\tau)}{\theta_1(u,\tau)}\left(\frac{\theta_{\sigma}(0,\tau)}{\eta}\right)^3,
\end{equation}
where $\theta_{\sigma} = \theta_{1,2,3,4}$ for $\sigma = \tilde{R}, R , NS , \tilde{NS}$ respectively and
$\epsilon = (1, (-1)^{w+1}, (-1)^{m+1}, (-1)^{w+m})$ for ($NS$, $\tilde{NS}$, $R$, $\tilde{R}$) respectively. This partition function includes all contributions from the worldsheet fermions and their spin structure and also the superconformal ghosts. To make this into a full string partition function, only a bosonic contribution should be added.\\

\noindent To make the link between these models, we rewrite the theta-functions by linking them directly as 
\begin{align}
\theta\left[
\begin{array}{c}
\frac{1}{2}+\frac{w}{N} \\
\frac{1}{2}+\frac{m}{N}  \end{array} 
\right] &= - e^{\pi i \tau \frac{w^2}{N^2}+\pi i \frac{w}{N} + \frac{2\pi i w m}{N^2}} \theta_1\left(\frac{w}{N}\tau + \frac{m}{N},\tau\right) , \\
\theta\left[
\begin{array}{c}
\frac{w}{N} \\
\frac{m}{N}  \end{array} 
\right] &= e^{\pi i \tau \frac{w^2}{N^2} + \frac{2\pi i w m}{N^2}}\theta_3\left(\frac{w}{N}\tau + \frac{m}{N},\tau\right), \\
\theta\left[
\begin{array}{c}
\frac{1}{2}+\frac{w}{N} \\
\frac{m}{N}  \end{array} 
\right] &= e^{\pi i \tau \frac{w^2}{N^2} + \frac{2\pi i w m}{N^2}}\theta_2\left(\frac{w}{N}\tau + \frac{m}{N},\tau\right), \\
\theta\left[
\begin{array}{c}
\frac{w}{N} \\
\frac{1}{2}+\frac{m}{N}  \end{array} 
\right] &= e^{\pi i \tau \frac{w^2}{N^2}+\pi i \frac{w}{N} + \frac{2\pi i w m}{N^2}}\theta_4\left(\frac{w}{N}\tau + \frac{m}{N},\tau\right).
\end{align}
The link between the $\sigma$-index and the $(\alpha, \beta)$ couple is:
\begin{align}
(1/2,1/2) &\to \tilde{R}, \\
(1/2,0) &\to R, \\
(0,1/2) &\to NS, \\
(0,0) &\to \tilde{NS}.
\end{align}
Again the saddle point can be handled quite easily. The saddle point integral yields again $1/k$, where all other factors present in the cigar partition function (\ref{cigarpt}) are simply to be evaluated at $s_1=\frac{w}{N}$ and $s_2 = \frac{m}{N}$. The rest is simply a bookkeeping exercise.\footnote{For the reader who is interested in more details, we make the following remarks. Starting with expression (\ref{flatpt}) and focussing on the holomorphic part with $\alpha$ and $\beta$, the sectors with $\beta = 1/2$ have a $e^{-\pi i w /N}$ factor in the $\omega'$'s which cancels with the $e^{\pi i w /N}$ phase present in the above conversion formulas. After this, all sectors above have the same prefactors, thus upon including the complex conjugate expression, only their modulus contributes and this gives a global prefactor $e^{-2\pi\tau_2 \frac{w^2}{N^2}}$, which cancels with the same prefactor appearing from the denominator.} The prefactor of $\left(\frac{1}{\left|\eta\right|^2\sqrt{4\pi^2\alpha'\tau_2}}\right)^{6}$ can be generated by including 8 free bosons and the bc ghosts, giving in total the contribution of 6 free bosons indeed.\footnote{Up to a $1/\sqrt{\tau_2}$ prefactor which in the notation of \cite{Lowe:1994ah} is absorbed into the fundamental domain measure.} \\

\noindent A question that immediately arises in this process is the following. For the flat orbifold, it is known that only odd $N$ makes sense as a string theory on a cone \cite{Dabholkar:1994ai}\cite{Lowe:1994ah}\cite{Adams:2001sv}. Yet on the cigar orbifold, no mention is made of such a restriction in the literature \cite{Sugawara:2012ag}\cite{Eguchi:2010cb}\cite{Sugawara:2011vg}. Although we should remark that in most of this work, the authors were interested in constructing consistent modular invariant partition functions (which is satisfied by the above expression also for even $N$). It seems then that also for these spaces, an interpretation in terms of strings on a cone can only be given for odd $N$. We postpone a deeper investigation into this issue to possible future work.\footnote{In fact, the argument given in \cite{Adams:2001sv} can be copied exactly for the cigar orbifolds. The orbifold identification has two possible actions on the spacetime spinors:
\begin{equation}
R = e^{\frac{2\pi i J}{N}} \quad \text{or} \quad R= (-)^Fe^{\frac{2\pi i J}{N}}
\end{equation}
where $J$ is the generator of angular rotations in the $U(1)$ cigar angular direction (which becomes $J_{89}$ on the plane of \cite{Adams:2001sv} by taking $k\to\infty$). Then we have $R^N = (-)^F$ or $R^N = (-)^{(N+1)F}$. The first possibility leads to an inclusion of $(-)^F$ in the orbifold group and the absence of spacetime spinors (which is unwanted). We hence should choose the second option with odd $N$ to avoid this.} \\

\noindent The untwisted sector ($w=m=0$) was handled in \cite{Giveon:2014hfa}. In more detail, it is given by
\begin{equation}
Z(\tau) \approx -\frac{k}{N\tau_2}\ln(\epsilon)\frac{2}{\pi}\frac{1}{4} \frac{\left|\theta_3^4 - \theta_4^4 - \theta_2^4\right|^2}{\left|\eta\right|^{12}},
\end{equation}
which equals the flat space cosmological constant (up to the factor of $1/N$) and it vanishes again due to Jacobi's obscure identity. Including the other flat dimensions, the $bc$-ghosts and the modular integral, we obtain
\begin{equation}
\label{superF}
Z \approx -\frac{1}{N}\int_{\mathcal{F}}\frac{d\tau d\bar{\tau}}{4\tau_2^2}k\ln(\epsilon)\frac{1}{2\pi} \left(\frac{1}{\left|\eta\right|^2\sqrt{4\pi^2\alpha'\tau_2}}\right)^{6} \frac{\left|\theta_3^4 - \theta_4^4 - \theta_2^4\right|^2}{\left|\eta\right|^{12}}.
\end{equation}
The transverse area can again be identified in this expression as the same formula (\ref{area}). \\ 

\begin{figure}[h]
\centering
\includegraphics[width=0.8\textwidth]{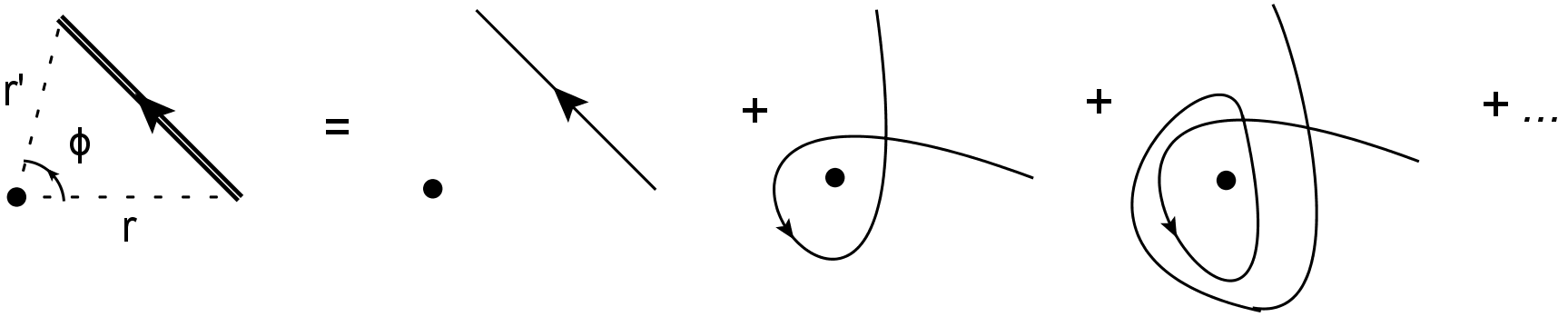}
\caption{The total Euclidean propagator in flat space between points $(r,0)$ and $(r',\phi)$ can be written as a sum over propagators with fixed winding number around the origin.}
\label{particlePolar}
\end{figure}

\noindent This seems a quite important result: even though the GSO projection assigns a special role to the angular coordinate, the partition function of Euclidean Rindler space is precisely the same as the flat space vacuum energy on an infinite 2d plane (just as it was for the bosonic string). \\
Let us end on a more speculative note here. This equality means the coordinate transformation from polar coordinates to cartesian coordinates is unhindered by the GSO projection. Besides being mathematically interesting, this has important physical consequences. \\
First let us go back to quantum field theory in curved spacetimes. We remind the reader that for quantum fields in Rindler space, the stress tensor vanishes in the Minkowski vacuum, which can be rewritten in terms of the coordinates of the Rindler observer as:
\begin{equation}
\left\langle T_{\mu\nu}\right\rangle_M = \left\langle T_{\mu\nu}\right\rangle_R + \text{Tr}_R\left(T_{\mu\nu}e^{-\beta H_R}\right)_{H_R \neq 0}.
\end{equation}
The thermal bath of Rindler particles combines with the Casimir contribution to give a vanishing vev. Thus the thermal bath does not backreact on the background. The ultimate reason for this is the fact that Rindler space and Minkowski spacetime are simply related by a coordinate transformation. \\
This QFT story can be interpreted in Euclidean signature as well. It was shown in \cite{Dowker:1977zj}\cite{Troost:1977dw}\cite{Troost:1978yk}\cite{Parentani:1989gq} that the Euclidean propagator in flat space can be expanded into a sum over winding numbers around the origin:
\begin{equation}
G(r,0;r',\phi; s) = \sum_{w\in\mathbb{Z}}G^{(w)}(r,0;r',\phi; s).
\end{equation}
This relation is shown diagrammatically in figure \ref{particlePolar}.
\\
The stress tensor vev can then be obtained by applying a suitable differential operator on the Green function and taking the coincidence limit of the two points. Hence we obtain
\begin{equation}
\left\langle T_{\mu\nu}\right\rangle_E = \left\langle T_{\mu\nu}\right\rangle_{w=0} + \sum_{w'=-\infty}^{+\infty}\left\langle T_{\mu\nu}\right\rangle_{w},
\end{equation}
where the prime denotes the absence of the $w=0$ term in the sum. The $w=0$ term is the temperature-independent Casimir contribution whereas the remaining sum is the thermal contribution. In this Euclidean setting, it is apparent that the vanishing of the vev in 2d flat space implies that the sum of the Casimir and the thermal contributions vanish. Again the main reason is the coordinate equivalence between cartesian coordinates and polar coordinates. \\
Now back to string theory. The fact that the partition function of string theory in polar coordinates is the same as that in cartesian coordinates shows that the coordinate transition between both is still valid in string theory. This is a necessary condition to have a vanishing stress tensor.\footnote{It seems difficult to make this point more firmly: the full stress tensor of the string gas seems difficult to obtain; we only obtained the most dominant contribution in \cite{Mertens:2014dia}. The free energy however is directly linked to the canonical internal energy which is to be interpreted as the spatial integral of $T_0^0$ and the result we obtain is then only in the weaker integrated sense.} In this sense, the tip of the cigar is not special and the GSO projection does not ruin the coordinate equivalence between polar and cartesian coordinates. This suggests there is no backreaction caused by the thermal atmosphere of the black hole (a good thing!). 

\subsection{Continuation of the flat orbifold inherited from the cigar orbifold}
Let us make a short detour here and consider a question first posed in \cite{Dabholkar:1994ai}: can we continue the partition functions on the $\mathbb{C}/\mathbb{Z}_N$ orbifolds to a non-integer $N$? An immediate response would be no, since modular invariance is no longer present for such values of $N$. Overcoming the initial shock, one might be tempted to use this approach as a possible off-shell proposal for string theory on a cone. In fact, we argued in \cite{Mertens:2013zya} that for the dominant behavior (given in field theory language) such a continuation is quite natural. Even if one believes this, the partition functions on $\mathbb{C}/\mathbb{Z}_N$ do not lend themselves towards continuation in $N$ (as was discussed by Dabholkar as well \cite{Dabholkar:1994ai}). Here, we perform this natural continuation at the level of the partition function on the cigar orbifold. Our goal is to use this continuation of the cigar CFT to tell us something about the continuation for the flat cones. \\
The partition function (\ref{starting}) on the cigar orbifold can be rewritten in the suggestive way
\begin{align}
Z &= \frac{1}{N}2\sqrt{k(k-2)}\int_{\mathcal{F}}\frac{d\tau d\bar{\tau}}{\tau_2} \int_{0}^{1}ds_1ds_2 \nonumber \\
&\sum_{m,w=-\infty}^{+\infty}\sum_i q^{h_i}\bar{q}^{\bar{h}_i}e^{4\pi\tau_2(1-\frac{1}{4(k-2)}) -\frac{k\pi}{\tau_2}\left|(s_1 - \frac{w}{N})\tau +(s_2 - \frac{m}{N})\right|^2+2\pi\tau_2s_1^2} \nonumber \\
&\frac{1}{\left|\sin(\pi(s_1\tau + s_2))\right|^2}\left|\prod_{r=1}^{+\infty}\frac{(1-e^{2\pi i r \tau})^2}{(1-e^{2\pi i r \tau - 2\pi i (s_1\tau +s_2)})(1-e^{2\pi i r \tau + 2\pi i (s_1\tau +s_2)})}\right|^2.
\end{align}
This allows a natural continuation in $N$ as $1/N \to \frac{\beta}{\beta_{\text{Hawking}}}$, where modular invariance is lost. We investigate how this translates into a continuation of the flat cone.\\
Firstly, the $s_1$- and $s_2$-integrals are only over a unit interval here. The untwisted sector has two stationary points for each $s$-integral: $s=0$ and $s=1$.\footnote{In fact, the stationary point $s_1=1$ corresponds to $w=1$ which gets translated upon using Poisson resummation etc. into the winding $1$ mode. Higher winding modes (which would correspond to $s_1=2$ etc.) are not stationary points. This shows from this perspective as well that only singly wound modes are present in the large $k$ limit. As a reminder, the $w$ and $m$ quantum numbers are the torus cycle winding numbers. The $m$ quantum number gets Poisson resummed into the discrete momentum whereas the winding number $w$ remains the same (at least for the discrete representations) throughout the manipulations.} However, both are at the boundary of the integration interval and hence receive weight factor $1/2$. Both contributions are equal due to the periodicity of the Ray-Singer torsion. In all, one can choose one of these saddle points and neglect the weight factors. This agrees with our earlier analysis of the saddle points.\\
Upon making the replacement for $1/N$, one finds a saddle point only for those $w$ and $m$ for which
\begin{equation}
\label{inequal}
0 \leq \frac{w\beta}{\beta_{\text{Hawking}}} \leq 1 , \quad 0 \leq \frac{m\beta}{\beta_{\text{Hawking}}} \leq 1
\end{equation}
holds. The lower boundary corresponds to the untwisted sector and the upper boundary is only reached precisely for the orbifold models. For such general values of $\beta$, the points $(s_1, s_2) = (1,0),\, (0,1),\, (1,1)$ are not saddle points anymore. This implies the untwisted sector has an overall scaling of $1/4$ with respect to the orbifold points. The only difference for the twisted sectors is hence the replacement of the twisted sum by
\begin{equation}
\frac{1}{N} \sum_{m,w=0}^{N-1} \to \frac{\beta}{\beta_{\text{Hawking}}}\sum_{m,w}'
\end{equation}
where the prime indicates that $m$ and $w$ are integers restricted by (\ref{inequal}). One readily checks that for the orbifold points, one regains the earlier results. \\
In particular, this continuation implies that for $T<T_{\text{Hawking}}$ the only sector present is the $m=w=0$ sector and this is unchanged as $\beta$ is varied. This is in conflict with the free-field trace $\text{Tr}e^{-\beta H}$ which decreases monotonically as $\beta$ increases. \\
We note that the most dominant thermal state ($w=\pm1, m =0$) is present as soon as $T>T_{\text{Hawking}}$. At the Hawking temperature itself, we saw earlier that this state is in fact also present, albeit camouflaged in the flat space result. Thus one can follow this state as one lowers the temperature all the way to the Hawking temperature where this state becomes marginal. This seems to fit with our general expectations on continuing this state through a range of temperatures. \\
Also, this continuation makes it clear that this resulting off-shell proposal is \emph{non-analytic} in $N$, in contrast to the arguments made by Dabholkar. \\
The arguments presented in this section should not be viewed as a full-fledged proposal for the off-shell continuation of the flat cones, we merely link the most natural off-shell continuation of the cigar orbifolds to the inherited off-shell continuation of the flat cones. Whether these continuations make sense, is left open. \\

\noindent What is apparent is that one gets another hint that these partition functions do not appear to correspond to Hamiltonian free field traces (other hints are the discussions made by Susskind and Uglum \cite{Susskind:1994sm} and the absence of certain winding sectors in the thermal spectrum (i.e. the unitarity constraints of the model), obscuring a thermal interpretation \cite{Mertens:2014cia}). \\

\noindent In the next section, we will make a concrete proposal explaining this discrepancy.

\section{String thermodynamics and modular domains for contractible thermal circles}
\label{modDom}
The discussions made in the previous section actually allow us to discuss more deeply the role of our starting point: does one define string thermodynamics in the fundamental domain or in the strip? Are these descriptions always identical? We shall first answer this question in the affirmative in a general background, after which we will uncover a puzzle with the precise partition functions discussed in the previous section. The discrepancy will be explained by a more detailed discussion on the different torus embeddings that are actually path integrated over.

\subsection{Generalization of the McClain-Roth-O'Brien-Tan theorem}
\label{Mcr}
The general torus path integral on the fundamental domain for a general (on-shell) background\footnote{For simplicity we write down only a metric background here, but the result is more general.}
\begin{equation}
\label{funda} 
Z_{T_2} = \int_{\mathcal{F}} \frac{d\tau_2}{2\tau_2} d\tau_1 \Delta_{FP} \int \left[\mathcal{D}X\right]\sqrt{G}
\exp -\frac{1}{4\pi\alpha'} \int d^2\sigma \sqrt h h^{\alpha \beta} \partial_\alpha X^\mu \partial_\beta X^\nu G_{\mu\nu}(X),
\end{equation}
where $\Delta_{FP}$ is the Faddeev-Popov determinant $\left|\eta\right|^4$ and with the torus boundary conditions (for some periodic field $X$)
\begin{align}
X(\sigma^1+2\pi,\sigma^2) & =  X(\sigma^1,\sigma^2) + 2\pi w R, \\
X(\sigma^1+2\pi\tau_1,\sigma^2+2\pi\tau_2) & =  X(\sigma^1,\sigma^2) + 2\pi m R,
\end{align}
can be rewritten in the strip domain as
\begin{equation}
\label{stri} 
Z_{T_2} = \int_0^\infty \frac{d\tau_2}{2\tau_2} \int_{-1/2}^{1/2} d\tau_1 \Delta_{FP} \int \left[\mathcal{D}X\right]\sqrt{G}
\exp -\frac{1}{4\pi\alpha'} \int d^2\sigma \sqrt h h^{\alpha \beta} \partial_\alpha X^\mu \partial_\beta X^\nu G_{\mu\nu}(X),
\end{equation}
with torus boundary conditions
\begin{align}
X(\sigma^1+2\pi,\sigma^2) & =  X(\sigma^1,\sigma^2), \\
X(\sigma^1+2\pi\tau_1,\sigma^2+2\pi\tau_2) & =  X(\sigma^1,\sigma^2) + 2\pi r R. 
\end{align}
In flat space, this equality was established by \cite{McClain:1986id}\cite{O'Brien:1987pn} some time ago. In their proof, the authors make explicit use of the flat space worldsheet action. It turns out (almost trivially) that one can make the argument independent of the flat space action and hence generalize it to an arbitrary conformal worldsheet model. \\
Starting in the fundamental domain, the proof uses that the effect of a modular transformation can be undone by a redefinition of the wrapping numbers:
\begin{align}
T&: m \to m+ w, \\
S&: m \to -w, \quad w \to m.
\end{align}
One does not need the precise action for this. What is required is that the worldsheet theory is conformally invariant. The proof then follows exactly the same strategy as for flat space: one can map each $(m, w)$ sector into $(r, 0)$ by a modular transformation, precisely building up the strip modular domain in the process. These steps are made with much more care in appendix \ref{MROB}. \\

\noindent Note that no use is made of the non-contractibility of the $X$-cycle: in fact one can apply this to a contractible circle as well (like the angular coordinate in polar coordinates). This shows in a very general way that the fundamental domain and the strip domain give equal results in any spacetime.

\subsection{What do these path integrals represent for Euclidean Rindler space?}
\begin{figure}[h]
\begin{minipage}{0.32\textwidth}
\centering
\includegraphics[width=0.9\textwidth]{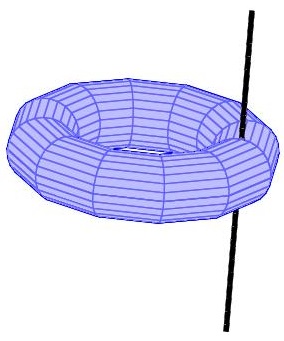}
\caption*{(a)}
\end{minipage}
\begin{minipage}{0.32\textwidth}
\centering
\includegraphics[width=0.9\textwidth]{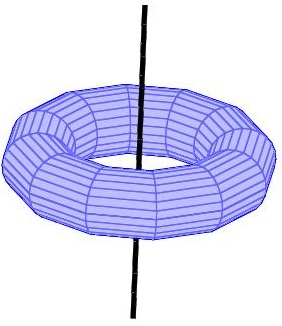}
\caption*{(b)}
\end{minipage}
\begin{minipage}{0.32\textwidth}
\centering
\includegraphics[width=0.99\textwidth]{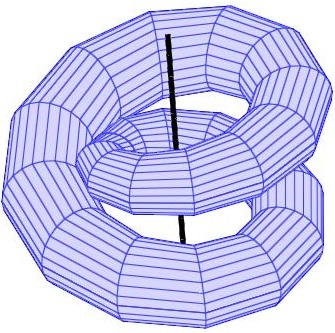}
\caption*{(c)}
\end{minipage}
\caption{(a) Singly wound torus that intersects the axis (perpendicular to the Rindler plane) through the Euclidean Rindler origin. Such configurations are not path integrated over in (\ref{stri}). (b) Singly wound torus that does not intersect the axis through the Euclidean Rindler origin. Such configurations are path integrated over in (\ref{stri}). (c) Twice wound torus that again does not intersect the axis through the origin. Such a configuration corresponds to a free closed string thermal trace.}
\label{torusPolar1}
\end{figure}

Now we come to an important point that we did not discuss at all in \cite{theory}. \\
The above configurations (\ref{funda}) and (\ref{stri}) do \emph{not} represent the most general embedding of the worldsheet torus into the target space, when this space has a contractible $X$-circle. It is most transparent to discuss this first in the strip domain. The string path integral with winding only along one of the torus cycles (\ref{stri}) represents a restricted set of tori embeddings in the target space when the thermal circle is contractible. Upon shifting the dependence on the moduli from the boundary conditions to the worldsheet metric, the torus wrapping (along only the temporal worldsheet direction) is imposed as
\begin{equation}
X^0(\sigma^1, \sigma^2+1) = X^0(\sigma^1,\sigma^2) + r \beta,
\end{equation}
where $\sigma^1$ is the spatial coordinate on the worldsheet and $\sigma^2$ is the timelike coordinate. The interpretation of this boundary condition is that \emph{all} points along a fixed $\sigma^2$-slice rotate along the (Euclidean) time dimension to form a 2-torus. As an example, let us take a closer look at Rindler space. For Euclidean Rindler space, this means that configurations such as \ref{torusPolar1}(a) are not integrated over: points in the ``inner'' path do not rotate around the Rindler origin. Configurations displayed in figure \ref{torusPolar1}(b) on the other hand are the ones that we take into account.

\noindent It is interesting to point out that the configurations shown in figure \ref{torusPolar1}(a) are associated in \cite{Susskind:1994sm} to open-closed interactions in the Lorentzian picture and these are entirely missed in the above path integral approach to string thermodynamics, suggesting that the non-interacting closed string trace only is contained in the above path integral. The higher winding numbers are associated to tori that wind around the Rindler origin multiple times, such as that displayed in the right most figure of figure \ref{torusPolar1}.

\noindent On the fundamental domain with 2 wrapping numbers (\ref{funda}), the same story happens: some torus embeddings are completely missed. Crucial in this argument is that the above torus boundary conditions do not allow mixed wrapping numbers (a property that could be allowed in a contractible target space), such as the torus embedding in the left figure of figure \ref{torusPolar1}. This means the Susskind-Uglum interactions are completely missed in the above path integrals, suggesting that they give only non-interacting (free-field) closed strings.\footnote{A point of critique is to be mentioned here: this path integral represents one string loop on the thermal manifold. Even in field theory, it is unclear whether the free field trace and the one-loop result on the thermal manifold agree for gauge fields (and for higher spin fields as well) \cite{Kabat:1995eq}\cite{Kabat:1995jq}\cite{Donnelly:2012st}\cite{Donnelly:2014fua}. For a very interesting recent development, see \cite{He:2014gva}. 
Presumably, this string path integral should hence be compared with the Hamiltonian trace over Lorentzian fields.} \\

\subsection{Explicit CFT result}
The other approach to Euclidean Rindler space (starting from the cigar CFT and then taking the small curvature (large $k$) limit), leads to apparently different results.
The bosonic partition function \cite{Hanany:2002ev} is given by
\begin{align}
\label{bossonic}
Z &= 2\sqrt{k(k-2)}\int_{\mathcal{F}}\frac{d\tau d\bar{\tau}}{\tau_2} \int_{0}^{1}ds_1ds_2 \nonumber \\
&\sum_{m,w=-\infty}^{+\infty}\sum_i q^{h_i}\bar{q}^{\bar{h}_i}e^{4\pi\tau_2(1-\frac{1}{4(k-2)}) -\frac{k\pi}{\tau_2}\left|(s_1 - w)\tau -(s_2 - m)\right|^2+2\pi\tau_2s_1^2} \nonumber \\
&\frac{1}{\left|\sin(\pi(s_1\tau - s_2))\right|^2}\left|\prod_{r=1}^{+\infty}\frac{(1-e^{2\pi i r \tau})^2}{(1-e^{2\pi i r \tau - 2\pi i (s_1\tau -s_2)})(1-e^{2\pi i r \tau + 2\pi i (s_1\tau -s_2)})}\right|^2.
\end{align}
As explicitly demonstrated above in equation (\ref{flatbosonic}), the large $k$ limit yields the partition function of an infinite 2d plane. \\
In \cite{Sugawara:2012ag}, Sugawara wrote down the partition function obtained upon setting $w=0$ and simultaneously extending the modular integration to the entire strip. This application of the McClain-Roth-O'Brien-Tan theorem is a bit naive as we demonstrate here. The candidate partition function equals
\begin{align}
Z_{\text{candidate}} &= Z_{m=w=0} + 2\sqrt{k(k-2)}\int_{E}\frac{d\tau d\bar{\tau}}{\tau_2} \int_{0}^{1}ds_1ds_2 \nonumber \\
&\sum_{m'=-\infty}^{+\infty}\sum_i q^{h_i}\bar{q}^{\bar{h}_i}e^{4\pi\tau_2(1-\frac{1}{4(k-2)}) -\frac{k\pi}{\tau_2}\left|s_1\tau -(s_2 - m)\right|^2+2\pi\tau_2s_1^2} \nonumber \\
&\frac{1}{\left|\sin(\pi(s_1\tau - s_2))\right|^2}\left|\prod_{r=1}^{+\infty}\frac{(1-e^{2\pi i r \tau})^2}{(1-e^{2\pi i r \tau - 2\pi i (s_1\tau -s_2)})(1-e^{2\pi i r \tau + 2\pi i (s_1\tau -s_2)})}\right|^2,
\end{align}
where the prime on the summation index denotes that we do not include the $m=0$ term. Taking the large $k$ limit, one finds (upon dropping the $(m,w) = (0,0)$ sector) almost the same expression as (\ref{flatbosonic}):\footnote{Upon reincluding the factor of $1/N$.}
\begin{equation}
Z_{\text{candidate}} \sim V_T \mathcal{A}\int_{E}\frac{d\tau d\bar{\tau}}{4\tau_2^2}\frac{1}{{\tau_2}^{12}}\left|\eta(\tau)\right|^{-48},
\end{equation}
the only difference being the change in modular integration domain. What one can say about this, is that the simple replacement of the strip $E$ with $\mathcal{F}$ and the inclusion of a second sum (over $w$) does \emph{not} give equal quantities. If it did, then the large $k$ limits should be the same as well.\footnote{It is instructive to follow the arguments in either of these papers \cite{McClain:1986id}\cite{O'Brien:1987pn} as far as possible. It turns out that the holonomy integrals over $s_1$ and $s_2$ in the end are causing the mismatch: these parameters transform as a doublet under modular transformations (just as $m$ and $w$), the problem then finally is the region of integration of these variables, which does not allow a clean extraction of the sum over different modular images to build up the strip domain.} \\
For type II superstrings, one would obtain instead (again dropping the $(m,w) = (0,0)$ sector)
\begin{equation}
\label{typeIIstri}
Z_{\text{candidate}} \sim V_T\mathcal{A}\int_{E}\frac{d\tau_1 d\tau_2}{2\tau_2^2}\frac{1}{\tau_2}\left(\frac{1}{\left|\eta\right|^2\sqrt{\tau_2}}\right)^{6} \frac{\left|\theta_3^4 - \theta_4^4 - \theta_2^4\right|^2}{\left|\eta\right|^{12}} = 0.
\end{equation}\\

\noindent We want to emphasize two points on these observations. Firstly, the fundamental domain and strip results of this partition function are \emph{not} equal, apparently in contradiction to the arguments given in the previous subsection \ref{Mcr}. Secondly, for type II superstrings in the strip, the partition functions (\ref{superF}) and (\ref{typeIIstri}) vanish, a property which is (nearly) impossible for a free-field thermal trace. \\

\noindent How then is this consistent with our discussion in the two previous subsections? To that effect, let us take a closer look at the path integral boundary conditions used to obtain equation (\ref{bossonic}) in the first place. In \cite{Hanany:2002ev}, coordinate transformations were made obscuring the wrapping numbers around the cigar. Indeed, the temporal (i.e. angular) coordinate $\phi$ ($\phi \sim \phi + 2\pi$) was transformed into a single-valued coordinate $v$ as\footnote{In this formula, $r$ is the radial coordinate and $\rho$ is related to the gauge field of the gauged WZW model. These coordinates are not relevant for our discussion here.}
\begin{equation}
v = \sinh(r/2) e^{i\phi}e^{i\rho}
\end{equation}
and the remainder of the derivation focused on this coordinate. However, no fixed wrapping numbers along $\phi$ were specified in advance, and in the end both tori with fixed wrapping numbers along both cycles (i.e. those that do \emph{not} intersect the origin) and those that are partially wrapped are all in principle considered: they all get mapped into the same $v$ coordinate. Only in the end one again (re)identifies the winding numbers. \\
Moreover, we have shown in the previous section \ref{flatlimit} that the flat limit indeed reproduces the \emph{entire} 2d plane partition function, meaning these partially wrapped torus configurations are indeed taken into account. These facts strongly suggest that indeed the partially wrapped tori are considered as well for these partition functions.
 
\subsection{Discussion}
We conclude that the partition function result in \cite{Hanany:2002ev} actually is the genus 1 result on the thermal manifold and includes torus embeddings with non-definite winding number. The free-field trace on the other hand corresponds to the path integrals (\ref{stri}) or (\ref{funda}) in which the temporal coordinate has a definite winding number and these expressions are not the same as the full genus 1 result. \\

\noindent To proceed then, imagine we focus first on manifolds where all winding modes are present on the thermal manifold (topologically stable thermal circles). In previous work \cite{theory}, we obtained the most dominant (random walk) contribution directly from the string path integral (\ref{stri}) with torus boundary conditions (\ref{stri}), thereby reducing the string theory on the modular strip to a particle theory of the thermal scalar. Alternatively, from the field theory (and CFT) point of view, the most dominant mode (the thermal scalar) on the thermal manifold (on the modular fundamental domain) can be used to arrive at the same dominant behavior. Both of these approaches should match for spaces with topologically stable thermal circles and we used this in \cite{theory} to identify possible corrections to the random walk picture and to obtain the Hagedorn temperature on such a manifold. \\
 
\noindent This determines the random walk corrections fully for \emph{any} manifold: it seems impossible to imagine what sort of local corrections could be added to the (non-interacting) thermal scalar action that vanish on all manifolds with topologically stable thermal circles but are in general non-zero on the others. \\

\noindent From this perspective, when the manifold is topologically unstable in the thermal direction (such as for black holes or Rindler space), one could follow the worldsheet (i.e. string path integral) derivation of the thermal scalar starting from equation (\ref{stri}) and observe that the thermal scalar still determines the critical behavior, \emph{irrespective of whether it is present in the thermal spectrum (on the fundamental domain) or not}. \\
\noindent In particular, all winding numbers are present in the path integral result of section \ref{Mcr} and we can use the thermal spectrum simply as a tool to extract the form of the thermal scalar action. After that, the path integral of section \ref{Mcr} with its angular wrapping number stands on its own and represents the contribution from non-interacting closed strings. \\
\noindent Such a scenario would also solve our understanding of the BTZ WZW models where thermal winding modes are simply absent \cite{Mertens:2014nca}. The free string path integral (\ref{stri}) on the other hand has no problem with non-zero thermal windings. This discrepancy hence seems to be again related to the special torus embeddings that in the end cause the absence of thermal winding modes. \\
\noindent We finally remark that this would imply that the bosonic non-interacting free energy for Rindler space diverges (due to \emph{all} windings since all of them are tachyonic (these are simply not present in the genus-1 result)). All winding (non-oscillator) modes are tachyonic and localized at a string length from the horizon (higher winding modes are localized even more closely to the event horizon). This seems to be in agreement with the maximal acceleration phenomenon of \cite{Parentani:1989gq}. \\

\noindent With this understanding of the two different ways of studying string thermodynamics, one can ask which one is the most natural? The path integral approach leads to non-interacting strings in a fixed background, whereas the thermal manifold (CFT) approach leads to the full genus 1 result, which includes interactions with open strings stuck on the horizon. These interactions are quite exotic, since for instance dialling down the string coupling $g_s$ does \emph{not} decouple in any way these interactions: they are inherent to the torus path integral on the thermal manifold. In summing over genera to obtain the full (perturbative) thermodynamics, they should be included as well. In the next sections this will be our state of mind. \\
Then what does the non-interacting trace do for us? It represents the sum over non-interacting closed strings and should correspond to the free-field trace, an object that can be constructed in principle as soon as the Lorentzian (non-interacting) string spectrum is known. The results of section \ref{Mcr} show that this quantity is modular invariant as well and hence has a well-defined meaning in string theory (as coming from a torus worldsheet).\footnote{It would be very interesting to examine this statement in more detail, but we will not attempt this here.} The difference between these two approaches can be illustrated by realizing that in fact the path integral approach of section \ref{Mcr} actually considers the perforated space where the fixpoint is removed from the space: one can arrive at this setting by for instance taking the pinching limit of a topologically supported thermal circle. String worldsheets are hence not allowed to cross this fixpoint. Reincluding this point in the geometry leads to the full genus one result, to be interpreted as both free strings and exotic open-closed interactions. \\

\noindent Finally let us remark that the fact that for the cigar CFT (and its flat limit) the singly wound string state is present in the thermal spectrum, implies that the dominant regime of both of these different partition functions (\ref{stri}) and (\ref{bossonic}) is actually the same: both are dominated by the thermal scalar. This implies the open-closed interactions are a subdominant effect near the Hagedorn temperature for Rindler space.

\section{Remarks on the string/black hole correspondence on the $SL(2,\mathbb{R})/U(1)$ cigar}
\label{remarks}

In this section, we are interested in the opposite limit as that studied up to this point: we comment on the behavior of the cigar CFT as we \emph{lower} $k$ to its critical value of $k=1$ or $k=3$ for type II and bosonic strings respectively. Note that this is an on-shell change of the black hole: no conical singularity is created during the process. It is known that several phenomena occur at the critical value of $k$ \cite{Giveon:2005mi} and in this section we look at this limiting process from the thermal scalar point of view. Qualitatively, one expects the string/black hole correspondence point \cite{Horowitz:1996nw} to occur as soon as the black hole membrane \cite{Thorne:1986iy} diverges to infinity \cite{Kutasov:2005rr}\cite{Giveon:2005jv}. This corresponds to the thermal scalar becoming non-normalizable. The $SL(2,\mathbb{R})/U(1)$ black hole provides an exact setting where the corresponding string/black hole phase transition can be better studied \cite{Giveon:2005mi}\cite{Parnachev:2005qr}. Even though the genus one partition function is incapable of seeing this transition directly (the transition is driven by non-perturbative effects), several suggestive clues will arise even at the one loop level. Moreover, we will see a continuous transition between a random walk corresponding to a Hamiltonian free-field trace, and a random walk coming from the discrete thermal scalar bound to the black hole horizon, providing some explicit evidence of the final remark presented in the previous section. \\ 
The string spectrum of the cigar CFT can be found in appendix \ref{cigspectr}. We remind the reader that for the discrete modes, the $SL(2,\mathbb{R})$ quantum number $j = M -l$ for $l=1,2,\hdots$ and $M$ is directly related to $m$. First of all, we note that the $l=1$ state for the type II superstring (the $l=2$ state for the bosonic case) are exactly marginal for all values of $k$ (if they exist) \cite{Kutasov:2000jp}. As we lower $k$, we encounter the special value of $k=3$. For this value of $k$ several things happen:
\begin{itemize}
\item{The asymptotic linear dilaton Hagedorn temperature becomes equal to the Hawking temperature.}
\item{The wavefunctions of the discrete states all become non-normalizable as can be seen by looking at the asymptotic linear dilaton behavior.}
\item{The one-loop thermal partition function does not include any discrete states anymore, in corroboration with the non-normalizability of the wavefunctions. In this setting, the states are absent since the contour shift used in \cite{Hanany:2002ev} does not cross poles anymore.}
\item{The lowest weight state in the continuum becomes marginal.}
\end{itemize}
Proofs of these statements can be found in appendix \ref{proof}. Exactly the same effects happen for the superstring at $k=1$. Of course, several of these features have been known from previous work \cite{Giveon:2005mi}\cite{Nakayama:2005pk}. What we want to emphasize in this section is how the one-loop free energy (and other thermodynamical quantities) behaves when approaching the critical $k$ value. As $k$ approaches this value, the thermal scalar wavefunction spreads without bound, until it disappears from the spectrum. Immediately thereafter, a continuous state takes over its role. 
During this procedure, the dominant part of the free energy (coming from the thermal scalar) changes in extensivity. For $k>3$, this part scales as the area of the horizon. For $k=3$, the mode is sensitive to the entire volume of the space. \\

\noindent It is nice to see in the explicit formulas that this critical value indeed occurs as soon as the continuous modes show Hagedorn behavior.\footnote{In fact, we believe this continuous process sheds some light on the nature of the thermal scalar on black hole horizons. We discussed earlier that the interpretation of the genus one partition function in terms of the Hamiltonian trace is obscured since possibly interactions are included with an open string gas on the horizon. We made this more explicit in section \ref{modDom} by pointing out a difference in our treatment of the string path integral and the genus 1 CFT result. This is not the case for the thermal scalar in the continuous representation here since this one is asymptotically precisely the same as the linear dilaton thermal scalar, yielding long string dominance for near-Hagedorn linear dilaton temperatures: this can be made explicit by splitting the free energy in two contributions: one coming from the horizon and one from the asymptotic linear dilaton space. The latter is insensitive to these subtleties about torus embeddings and corresponds to non-interacting strings \cite{Kutasov:2000jp}\cite{Barbon:2007za}. The process of lowering $k$ results in a continuous transition between dominance by the discrete thermal scalar living close to the horizon and the asymptotic continuous thermal scalar giving long string dominance in a free-field trace of the asymptotic space. The nature of these states appears to be the same and the free-field trace is expected to be sensitive to the same thermal scalar and critical temperature as the genus one partition function (at least for this CFT), something we argued for at the end of the previous section.} \\

\noindent We can now give a full picture of thermodynamics on the $SL(2,\mathbb{R})/U(1)$ cigar (or the near-extremal NS5 brane solution) as one varies $k$. \\
It was shown by Sugawara \cite{Sugawara:2012ag} that the thermal partition function can be written as a sum of two contributions:
\begin{equation}
Z = Z_{\text{fin}} + Z_{\text{asympt}},
\end{equation}
a finite part coming from the tip of the cigar (containing both the discrete modes and part of the continuum\footnote{This continuum part comes from the phase shift contribution in the density of states (or equivalently the reflection amplitude of the gravitational potential).}) and an asymptotic part which diverges due to the infinite volume available and which is precisely equal to the linear dilaton thermal partition function. This translates into $F = F_{A} + F_{V}$, where $F_A$ scales like the transverse area of the black hole and $F_V$ scales as the volume of the entire space.\\
\noindent Let us start by taking $k$ very large. In this case, the background approaches Rindler space and the free energy ($\propto Z$) is dominated by the thermal scalar. The thermodynamical quantities have a random walk interpretation and are given by a localized contribution at string length from the horizon. They have a part proportional to the volume (subdominant) and a part proportional to the black hole area. It is the latter part that should be interpreted as intrinsic to the black hole itself. \\
Lowering $k$ invalidates the Rindler approximation. Nonetheless, roughly the same story is valid when considering the thermal scalar wavefunction on the entire cigar geometry. The random walk is spreading more and more from the tip of the cigar. Thermodynamic quantities still satisfy the same scaling properties and the dominant contribution comes from the discrete thermal scalar bound to the (Euclidean) horizon. \\
As soon as $k$ reaches 1 (type II) or 3 (bosonic), the (discrete state) thermal scalar drops out from the spectrum. When nearing this value of $k$, the random walk spreads without bound over the entire cigar and becomes non-normalizable. At the same time, the asymptotically linear dilaton thermal scalar (coming from the continuous part of $Z$) becomes massless. The random walk of the latter looks asymptotically like the one studied in \cite{Mertens:2014cia} for the linear dilaton background. Not just the thermal scalar, but in fact all bound states disappear from the spectrum at this point. The bound states are a typical feature of string theory since they only exist because of the possibility of including winding numbers along the cigar, a possibility that is absent for field theory. At $k=3$ (or $k=1$), only the continuous spectrum remains. \\

\noindent It is also known that for this value of $k$, the black hole itself becomes non-normalizable \cite{Giveon:2005mi}. Also, infalling D0-branes change their radiation pattern \cite{Nakayama:2005pk}: normally most radiation is incident on the black hole. When $k$ reaches the critical value, a significant part of the radiation propagates outwards, contrary to our intuition about black holes. 
This process of lowering $k$ is sketched in figure \ref{rw} below.
\begin{figure}[h]
\centering
\includegraphics[width=0.7\textwidth]{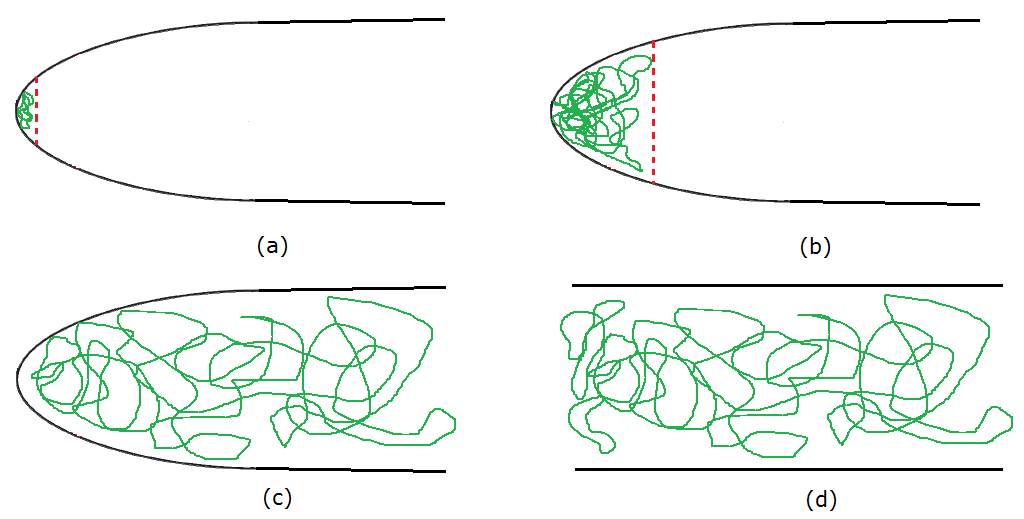}
\caption{(a) Large $k$ cigar background with random walk superimposed. Of course, in reality the random walk is on the spatial submanifold only, but for illustrative purposes we draw it like this. The (red) dashed line limits the spread of the discrete thermal scalar in the radial direction. For very large $k$ the flat Rindler approximation becomes accurate. (b) Lowering $k$ causes the discrete mode to spread radially. This invalidates the Rindler approximation. Nonetheless, the mode retains its dominant character. (c) When $k=1$, the mode spreads without bound radially. The mode becomes non-normalizable and disappears from the spectrum. Precisely at this point, the continuous thermal scalar takes over the dominant behavior. (d) The same dominant random walk on the linear dilaton background. The asymptotic parts of the random walks look the same for $k=1$.}
\label{rw}
\end{figure}

\section{Higher genus partition functions and the thermal scalar}
\label{higher}
In this section we consider the influence of the higher genera on the critical behavior. In \cite{Brigante:2007jv} arguments were given using dual field theories and worldsheet factorization that the Hagedorn temperature in general theories is non-limiting and that string perturbation theory (in $g_s$) actually breaks down near this temperature. It is of course interesting to try to apply some of these arguments to the case of black hole horizons. Similarly to \cite{Brigante:2007jv}, we will only focus on purely compact spaces for which the spectrum of the operator $\hat{\mathcal{O}}$ (as defined in (\ref{operatorO})) is entirely discrete. In the first subsection we will analyze the influence of a fixed higher genus contribution and in the second subsection we will consider the effect of summing over the higher genus corrections. We follow \cite{Brigante:2007jv} quite closely.
\subsection{Higher genus partition functions}
In this subsection we focus on the thermal partition function at fixed genus. Consider the $SL(2,\mathbb{R})/U(1)$ CFT (and its flat limit $k\to\infty$). We follow the decomposition of higher genus Riemann surfaces into propagators and 3-punctured spheres \cite{Polchinski:1988jq}. In general, higher genus Riemann surfaces (can) have divergences when various cycles pinch. The moduli of the propagators are sufficient to give the contribution from the boundary of moduli space, which is what we need. In case of topologically stable thermal circles, it was shown \cite{Brigante:2007jv} that winding conservation prohibits all propagators from containing the thermal scalar. For cigar-shaped thermal circles however, winding conservation is violated in scattering amplitudes \cite{Giveon:1999px}\cite{Giveon:1999tq}. \\
In \cite{Giveon:1999px}\cite{Giveon:1999tq} it was shown that $n$-point amplitudes can violate winding conservation by up to $n-2$ units, thus an amplitude such as $\left\langle V_{w=+1} V_{w=+1}V_{w=+1}\right\rangle$ or $\left\langle V_{w=-1} V_{w=-1}V_{w=-1}\right\rangle$ vanishes whereas $\left\langle V_{w=+1} V_{w=+1}V_{w=-1}\right\rangle$ or $\left\langle V_{w=+1} V_{w=-1}V_{w=-1}\right\rangle$ do not vanish and provide the relevant 3-point amplitudes; a property which persists in the large $k$ limit. Hence it \emph{is} possible in this case that all propagators contain the thermal scalar, simplifying the analysis.\footnote{One readily shows that this is possible without having to resort to giving any 3-point vertex a winding number violation by more than one unit.}
An explicit computation of this 3-string scattering amplitude will not be considered here, but we make some further remarks on these in appendix \ref{nnversusn}. In what follows we will describe this decomposition in much more detail.\\

\begin{figure}[h]
\centering
\includegraphics[width=0.8\textwidth]{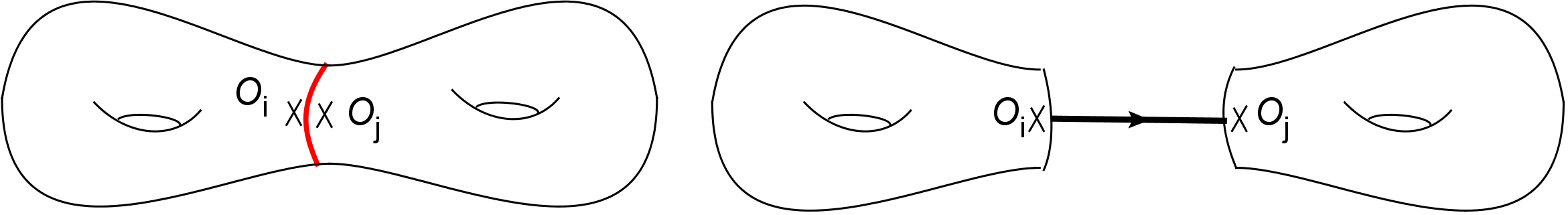}
\caption{Cutting open of a genus two worldsheet. For the cutting displayed here, this reduces it to two tori with single punctures and an intermediate propagator.}
\label{toruscutting}
\end{figure}

\noindent Following Polchinski \cite{Polchinski:1988jq}, we cut open the worldsheet (path integral) along a cycle and insert a complete set of states of local operators. 
Such a set of local operators have the normalization on the 2-sphere:
\begin{equation}
\left\langle O_i(\infty) O_j(0)\right\rangle_{S_2} \propto \delta_{h_i h_j}.
\end{equation}
We do not keep track of the overall normalization present in the Zamolodchikov metric written above.\\ 
We then have ($g=g_1+g_2$)\footnote{The index $j$ is restricted by $h_j = h_i$.}
\begin{equation}
\left\langle 1 \right\rangle_{g} = \int_{\left|q\right|<1} \frac{d^2q}{q\bar{q}}\sum_{i,j}q^{h_i}\bar{q}^{\bar{h}_i}\left\langle O_i(z_1) \right\rangle_{g_1}\left\langle O_j(z_2)\right\rangle_{g_2},
\end{equation}
where a new modulus $q$ is introduced (the sewing parameter), which should be integrated over a unit disk.\footnote{The alternative way of cutting the CFT reduces the genus by 1 as:
\begin{equation}
\left\langle 1 \right\rangle_{g} = \int_{\left|q\right|<1} \frac{d^2q}{q\bar{q}}\sum_{i,j}q^{h_i}\bar{q}^{\bar{h}_i}\left\langle O_i(z_1) O_j(z_2)\right\rangle_{g-1}.
\end{equation}
This way of cutting the worldsheet should be used as well in reducing the general genus $g$ amplitude and one readily adapts the formulas to incorporate also this procedure.} The intermediate set of states (labeled by $i$ (and $j$)) come in a propagator contribution, as can be seen by explicitly integrating over the modulus $q$. The cutting of the surface is demonstrated in figure \ref{toruscutting}.
The limit $q\to0$ corresponds to the pinching limit. In this limit, the lowest conformal weight dominates in the same way as that the lowest states dominate the one-loop partition function in the $\tau_2 \to \infty$ limit. We are interested in the most dominant contribution that is temperature-dependent, implying that the thermal scalar only is present in the intermediate vertex operators. The non-thermal massless operators are not temperature-dependent even though their conformal weight becomes equal to that of the thermal scalar at $\beta=\beta_H$. In the present section, we suppose that the wave operator $\hat{\mathcal{O}}$ of the thermal scalar (defined in equation (\ref{operatorO})) has a discrete spectrum $\psi_n$. Integrating over $q$ and keeping only the dominant contribution, this yields:
\begin{equation}
\left\langle 1 \right\rangle_{g} \approx \frac{1}{h_0+\bar{h}_0}\left\langle T_0(z_1) \right\rangle_{g_1}\left\langle T_0(z_2)\right\rangle_{g_2}
\end{equation}
where $0$ labels the state with lowest conformal weight of the thermal scalar $T$; this can be done cleanly due to the discreteness of the spectrum of $\hat{\mathcal{O}}$. We can rewrite it as
\begin{equation}
\left\langle 1 \right\rangle_{g} \approx \left\langle T_0 \right|\frac{1}{L_0+\bar{L}_0} \left| T_0 \right\rangle \left\langle T_0(z_1) \right\rangle_{g_1}\left\langle T_0(z_2)\right\rangle_{g_2}
\end{equation}
for the thermal scalar state $T_0$. Cutting along several cycles finally reduces the higher genus amplitude to a set of propagators and spheres with 3 punctures. One of the possible outcomes for the genus 2 worldsheet is shown in figure \ref{cuttingfull} below. 
\begin{figure}[h]
\centering
\includegraphics[width=0.4\textwidth]{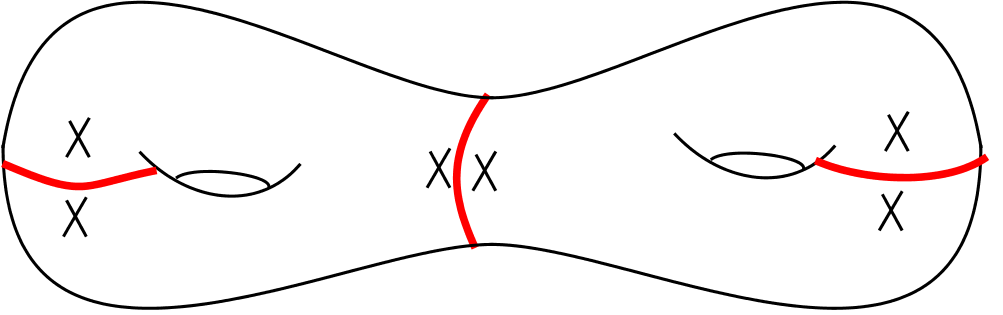}
\caption{Example of a full cutting of the genus two worldsheet into three propagators and two 3-punctured spheres.}
\label{cuttingfull}
\end{figure}
The 3-punctured spheres we need are the amplitudes of tree-level 3-thermal scalar scattering, which can in principle be computed.\\

\begin{figure}[h]
\centering
\includegraphics[width=5cm]{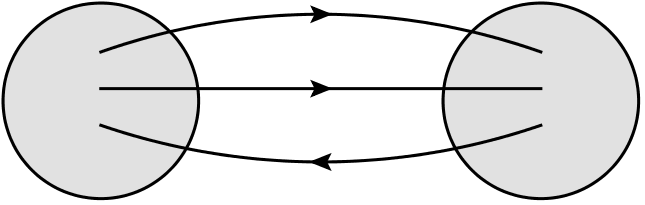}
\caption{Degenerate limit of the genus two surface in which the thermal scalar propagates along each internal line.}
\label{genustwo}
\end{figure}

\noindent Consider as an example the degenerate limit of a genus-2 amplitude shown in figure \ref{genustwo}.
Each propagator is dominated by the thermal scalar and behaves as $\sim 1/\lambda_0$. In string language we obtain for this vacuum amplitude $\mathcal{A}$:
\begin{equation}
\label{vacugen2}
\mathcal{A} = \frac{g_s^2}{\lambda_0^3}\left\langle T_0 T_0 T_0^* \right\rangle_{S^2}\left\langle T_0 T_0^* T_0^* \right\rangle_{S^2}.
\end{equation}
Next we would like to obtain a spacetime interpretation of these 3-vertex interactions. \\

\begin{figure}[h]
\centering
\includegraphics[width=3cm]{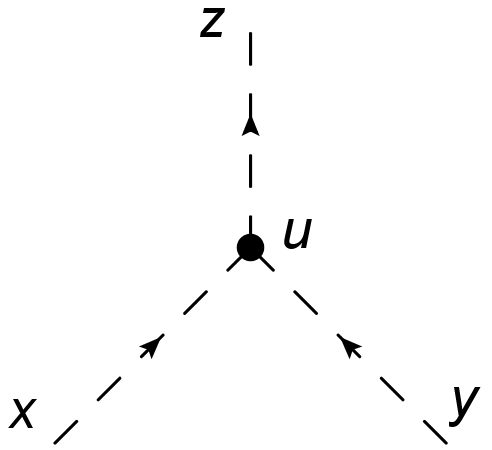}
\caption{Three-point amplitude in field theory.}
\label{threevertex}
\end{figure}

\noindent To that effect, let us consider the following off-shell 3-point function (figure \ref{threevertex}) in the field theory of the thermal scalar in coordinate space. For simplicity we are assuming that the three-vertex has amplitude $\lambda$ and does not include derivative interactions. Comments on these issues are provided further on.
With $\Delta(\mathbf{u},\mathbf{x})$ denoting the coordinate space scalar propagator, this amplitude equals
\begin{align}
\mathcal{A}_{\mathbf{x}\mathbf{y}\mathbf{z}} &= g_s\lambda \int d\mathbf{u} \Delta(\mathbf{u},\mathbf{x}) \Delta(\mathbf{u},\mathbf{y}) \Delta(\mathbf{z},\mathbf{u}) \nonumber \\
&= g_s\lambda \int d\mathbf{u} \sum_{m,n,o} \frac{\psi_m(\mathbf{u}) \psi_m(\mathbf{x})^*}{\lambda_m}\frac{\psi_n(\mathbf{u}) \psi_n(\mathbf{y})^*}{\lambda_n} \frac{\psi_o(\mathbf{z}) \psi_o(\mathbf{u})^*}{\lambda_o}.
\end{align}
It is clear from this expression that the amplitude gets its main support from that region in space where the wavefunctions live. \\
Instead focussing on external states with fixed quantum numbers ($l$, $r$, $s$), we would multiply this expression by the suitable wavefunctions:
\begin{equation}
\mathcal{A}_{lrs} = \int d\mathbf{x} \psi_l(\mathbf{x}) \int d\mathbf{y} \psi_r(\mathbf{y}) \int d\mathbf{z} \psi_s(\mathbf{z})^* \mathcal{A}_{\mathbf{x}\mathbf{y}\mathbf{z}}.
\end{equation}
This is the analog of Fourier transforming to momentum space in a translationally invariant spacetime. We obtain
\begin{align}
\mathcal{A}_{lrs} = g_s\lambda \int d\mathbf{u} \frac{\psi_l(\mathbf{u})}{\lambda_l}\frac{\psi_r(\mathbf{u})}{\lambda_r} \frac{\psi_s(\mathbf{u})^*}{\lambda_s}.
\end{align}
The propagator denominators are amputated to obtain the S-matrix elements and the $\mathbf{u}$-integral is the analog of $\delta(\sum k)$ in a translationally invariant space. From this expression, one sees that the amplitude $\mathcal{A}_{lrs}$ is mainly supported at the locations where the wavefunctions at the interaction location are relatively large. \\
In string theory, the three-vertex interaction (on the 2-sphere) can then be written in terms of the above field theoretical formulas as
\begin{equation}
\left\langle T_0 T_0 T^*_0 \right\rangle_{S^2} = \lambda \int d\mathbf{u} \psi_0(\mathbf{u})\psi_0(\mathbf{u}) \psi_0(\mathbf{u})^*.
\end{equation}
Thus the stringy amplitude determines $\lambda$ by stripping away the $\mathbf{u}$-integral given above.\\
Having gained insight into the space dependence of the 3-point interactions, we return to the genus 2 example given in equation (\ref{vacugen2}) and we obtain:
\begin{equation}
\label{3ptampl}
\mathcal{A} = \frac{g_s^2}{\lambda_0^3} \lambda^2 \int d\mathbf{u} \psi_0(\mathbf{u})\psi_0(\mathbf{u})\psi_0(\mathbf{u})^* \int d\mathbf{v} \psi_0(\mathbf{v})\psi_0(\mathbf{v})^*\psi_0(\mathbf{v})^*,
\end{equation}
We should remark that in principle the field theoretic coupling constant $\lambda$ will depend on the quantum numbers of the field factors present in the interaction vertex in the Lagrangian, but this does not cause any real difficulties.\footnote{In flat space, dependence of $\lambda$ on the momentum quantum number is interpreted as a derivative interaction. In general, such a derivative is not diagonal in the basis $\psi_l$. One should presumably sum an entire array of such derivative terms to obtain a diagonal quantity. The Lagrangian description in coordinate space is hence not particularly useful. Writing the Lagrangian in coordinates diagonal in $\psi_l$ is much better, but obscures the spacetime interpretation. 
Let us be more explicit about this point. A general 3-point interaction in field theory contains three field factors with some differential operators $\hat{D}_1$, $\hat{D}_2$ and $\hat{D}_3$ acting on them. Expanding the fields in a complete set of eigenmodes of $\hat{\mathcal{O}}$, one gets:
\begin{equation}
\mathcal{L} \supset g_s\lambda \int d\mathbf{x} \hat{D}_1 T(\mathbf{x}) \hat{D}_2 T(\mathbf{x}) \hat{D}_3 T^*(x) = g_s\lambda \sum_{l,m,n} a_l a_m a^*_n \int d\mathbf{x} \hat{D}_1 \psi_l(\mathbf{x}) \hat{D}_2 \psi_m(\mathbf{x}) \hat{D}_3 \psi_n^*(\mathbf{x}),
\end{equation}
where the c-numbers $a_n$ represent the modes that should be integrated in the path integral. This basis is diagonal in the relevant quantum numbers. The identification with string theory then proceeds by identifying field theory and string amplitudes as
\begin{equation}
\lambda \int d\mathbf{x} \hat{D}_1 \psi_l(\mathbf{x}) \hat{D}_2 \psi_m(\mathbf{x}) \hat{D}_3 \psi_n^*(\mathbf{x}) = \left\langle T_l T_m T^*_n \right\rangle_{S^2},
\end{equation}
which allows an identification of the differential operators and of $\lambda$ (which is now independent of the quantum numbers $l$, $m$ and $n$). This allows us to rewrite the interaction term in the field theory Lagrangian as
\begin{equation}
\mathcal{L} \supset g_s\lambda \int d\mathbf{x} \hat{D}_1 T(\mathbf{x}) \hat{D}_2 T(\mathbf{x}) \hat{D}_3 T^*(x) = g_s\sum_{l,m,n} \left\langle T_l T_m T^*_n \right\rangle_{S^2} a_l a_m a^*_n.
\end{equation}
Finally, one should modify equation (\ref{3ptampl}) into
\begin{equation}
\mathcal{A} = \frac{g_s^2}{\lambda_0^3} \lambda^2 \int d\mathbf{u} \hat{D}_1\psi_0(\mathbf{u})\hat{D}_2\psi_0(\mathbf{u})\hat{D}_3\psi_0(\mathbf{u})^* \int d\mathbf{v} \hat{D}_1^*\psi_0(\mathbf{v})^*\hat{D}_2^*\psi_0(\mathbf{v})^*\hat{D}_3^*\psi_0(\mathbf{v}).
\end{equation}
} \\

\begin{figure}[h]
\begin{minipage}{0.5\textwidth}
\centering
\includegraphics[width=0.75\textwidth]{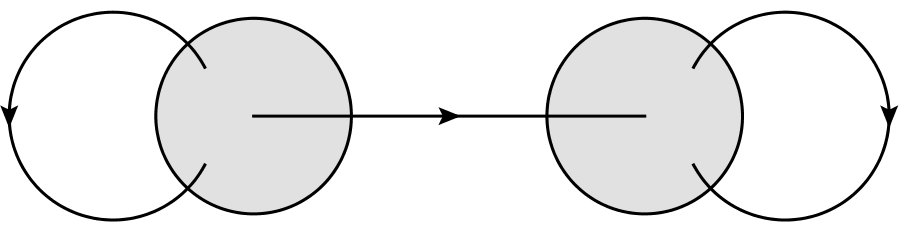}
\caption*{(a)}
\end{minipage}
\begin{minipage}{0.5\textwidth}
\centering
\includegraphics[width=0.75\textwidth]{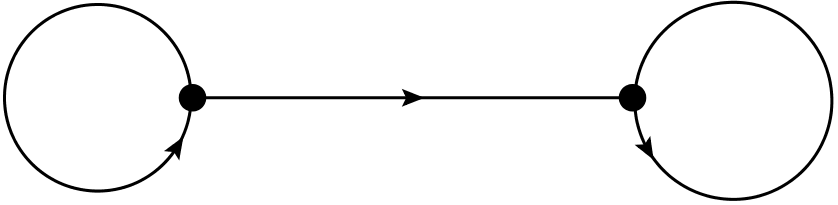}
\caption*{(b)}
\end{minipage}
\caption{(a) Other degenerate limit of the genus two surface. (b) Point particle graph of the alternate degenerate limit of the genus two surface.}
\label{alt1}
\end{figure}

\noindent This amplitude can now be rewritten as
\begin{align}
\mathcal{A} &= g_s^2 \lambda^2 \iint d\mathbf{u} d\mathbf{v} \frac{\psi_0(\mathbf{u})\psi_0(\mathbf{v})^*}{\lambda_0}\frac{\psi_0(\mathbf{u})\psi_0(\mathbf{v})^*}{\lambda_0}\frac{\psi_0(\mathbf{u})^*\psi_0(\mathbf{v})}{\lambda_0} \\
\label{second}
&\approx g_s^2 \lambda^2 \iint d\mathbf{u} d\mathbf{v} \left\langle \mathbf{u}\right| \frac{1}{L_0 + \bar{L}_0}\left| \mathbf{v}\right\rangle \left\langle \mathbf{u}\right| \frac{1}{L_0 + \bar{L}_0}\left| \mathbf{v}\right\rangle \left\langle \mathbf{v}\right| \frac{1}{L_0 + \bar{L}_0}\left| \mathbf{u}\right\rangle.
\end{align}
where in the second line, one should focus on the most dominant mode of the thermal scalar (hence the approximation symbol).\footnote{Again suitable insertions of the differential operators $\hat{D}_1$, $\hat{D}_2$, $\hat{D}_3$ should be made whenever necessary.} \\
Both of these equalities teach us something about the higher genus amplitudes. \\
The first line makes it clear that the amplitude gets its contribution from the near-horizon region (since that is where the thermal scalar wavefunction is supported). It shows that higher genus amplitudes have in their dominating contribution only support close to the tip of the cigar. This implies that even higher loop corrections cannot modify the fact that the random walk has a string-scale spread from the horizon.\footnote{Note that for instance in flat space, higher genus contributions would oscillate all over space, much like the genus one contribution. In a microcanonical scene, one expects these corrections to contract the long string into a stringy ball.} This random walk will have self-intersections (according to the above 3-point vertex), but it will still remain close to the horizon. We are led to the conclusion that Susskind's picture even holds when including higher loop interactions. \\
The second line (\ref{second}) in the above formula is in the first-quantized particle language, where the amplitude has the meaning of the following figure \ref{loop}.
\begin{figure}[h]
\centering
\includegraphics[width=4cm]{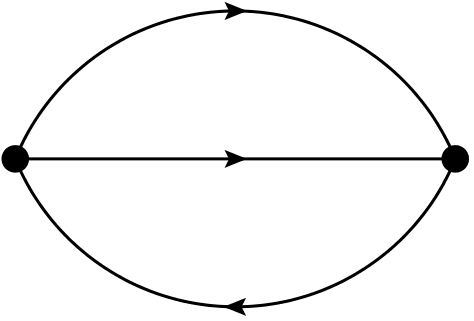}
\caption{First quantized Feynman graph corresponding to the above degenerate limit of the genus two surface.}
\label{loop}
\end{figure}
The spatial shape of the first-quantized thermal scalar should be interpreted as the shape of the long string. Note that the other possible degenerate limit of the genus-2 surface yields the same most dominant contribution (\ref{3ptampl}) as shown in figure \ref{alt1}.

\noindent The random walk picture is immediately distilled from this in a very explicit manner by using the Schwinger trick on the propagators:\footnote{In this formula, the ghost contribution is included in $L_0$ and $\bar{L}_0$.} 
\begin{equation}
\frac{1}{L_0 + \bar{L}_0} = \int_{0}^{+\infty}dT e^{-(L_0 + \bar{L}_0)T}
\end{equation}
and then giving a Lagrangian interpretation to this Hamiltonian picture amplitude. In this picture, each propagator contains a proper time parameter (Schwinger parameter). What one finds is that there is a set of open random walks with 3-point intersections. The locations of these interactions are also integrated over the entire space. The amplitude receives the largest contribution from the near-horizon region. Note that the genus $g$ diagram gives rise to a $g$-loop random walk. \\

\begin{figure}[h]
\begin{minipage}{0.5\textwidth}
\centering
\includegraphics[width=0.4\textwidth]{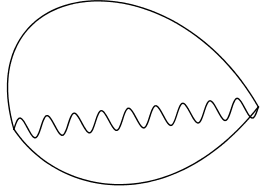}
\caption*{(a)}
\end{minipage}
\begin{minipage}{0.5\textwidth}
\centering
\includegraphics[width=0.3\textwidth]{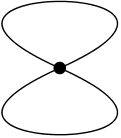}
\caption*{(b)}
\end{minipage}
\caption{(a) Graviton exchange by the long string. (b) Effective vertex of graviton exchange by integrating out virtual particles.}
\label{grav1}
\end{figure}

\noindent Also note that for instance graviton exchange by the long string is a subdominant effect (compared to other amplitudes at the same genus) for topologically trivial thermal circles: it has two separate 3-point interactions with a virtual graviton in between. This gets translated to an effective 4-point vertex, which dominates if winding number is conserved in interactions, which it is not in our case. This is shown in figure \ref{grav1}.

\noindent For topologically non-trivial thermal circles, the lowest interaction vertices are 4-point vertices, implying a different random walk intersection behavior. Such a 4-vertex arises by summing over all possible internal lines between two 3-vertices. 3-vertices on the other hand can only be associated to a pointlike interaction between three fields. \\

\noindent Of course, this reduction to particle path integrals, reintroduces UV divergences (at higher genera manifested as point-vertices), although we know this is merely an artifact of our approach. \\

\noindent For type II superstrings, we expect this picture to remain intact: the 3-punctured spheres would have different values, the Hagedorn temperature would be different, but the strategy and result would be the same. \\

\noindent As a summary, we conclude that at any \emph{fixed} genus, the partition function is dominated by the thermal scalar. The partition function behaves as a random walk with 3-point interactions and is localized close to the black hole event horizon. Note though that the propagator denominators actually force the partition function to diverge at $\beta = \beta_{\text{Hawking}}$ (as for the one-loop amplitude). \\
Of course, in string theory the different genera partition functions should be summed over and we now turn to this question. 

\subsection{The perturbative genus expansion and its limitations}
The single string partition function for topologically trivial thermal circles has the leading form (as determined above):
\begin{equation}
\label{expan}
Z_1 = - \ln(\beta-\beta_H) + \sum_{n=1}^{+\infty}C_n\frac{g_s^{2n}}{(\beta-\beta_H)^{3n}}.
\end{equation}
In this case, for each genus only the leading divergence is kept. This result can be found by using a double scaling limit as in \cite{Brigante:2007jv} where $g_s \to0$ and $\beta \to \beta_H$ while keeping $g_s^2 \propto (\beta-\beta_H)^3$. This limit ensures all other divergences at each genus scale out. \\
This expression can be seen to come directly from a field theory action. Consider the thermal scalar field theory action:\footnote{The $\hat{D}$ operators are differential operators that have been defined in a footnote earlier. $cc$ denotes the complex conjugate term of the second term in this expression.}
\begin{equation}
S = \int dV\sqrt{G}e^{-2\Phi}\left[T^* \hat{\mathcal{O}} T  + g_s \lambda \hat{D}_1 T\hat{D}_2 T \hat{D}_3 T^* +(cc)\right].
\end{equation}
In expanding the field $T$ in a complete set of eigenfunctions of $\hat{\mathcal{O}}$ with canonical normalization, one only retains the lowest mode in the critical limit:
\begin{equation}
T(\mathbf{x}) = \sum_n a_n \psi_n(\mathbf{x}) \approx a_0 \psi_0(\mathbf{x}).
\end{equation}
The action then reduces to
\begin{equation}
S \approx \lambda_0 a_0 a_0^*  + g_s \left\langle T_0 T_0 T_0^* \right\rangle_{S^2} a_0 a_0 a_0^* + (cc).
\end{equation}
The resulting critical diagrams then originate from the following theory:
\begin{equation}
Z_1 = \log \int d\phi d\phi^* e^{-\lambda_0 \phi \phi^* - g_s\phi\phi^*\left(\tilde{\lambda} \phi + \tilde{\lambda}^*\phi^*\right)},
\end{equation}
where $\phi = a_0$ is a complex number and the coupling $\tilde{\lambda} = \left\langle T_0 T_0 T_0^* \right\rangle_{S^2}$. The lowest eigenvalue $\lambda_0 \sim \beta-\beta_H$. With these definitions, one readily finds explicitly the expansion
\begin{equation}
Z_1 = -\ln\lambda_0 + \frac{6g_s^2\left|\lambda\right|^2}{\lambda_0^3} + \frac{162g_s^4\left|\lambda\right|^4}{\lambda_0^6} + \hdots
\end{equation}
agreeing with the previous expansion (\ref{expan}) and concretely giving values for the coefficients. \\

\noindent Clearly taking $\lambda_0 \to 0$ results in an infinite partition function: higher order terms are needed to determine the full (interacting) thermal scalar action. \\

\noindent If this scaling limit is not followed, additional contributions should be added to the functional integral (corresponding to the subleading corrections at each genus). These introduce 4-point interactions (and higher). As it stands, the above integral is non-perturbatively defined (although infinite in the absence of higher order corrections). Just like in \cite{Brigante:2007jv}, the thermal scalar potential can in principle be determined in this way.\\

\noindent The punch line is that the string perturbation series (the genus expansion) is not good in this case. In \cite{Brigante:2007jv} this was argued for for topologically stable thermal circles, where the perturbation series was seen to break down near the Hagedorn temperature. In this case however, since $T_H = T_{\text{Hawking}}$, we conclude that \emph{string perturbation theory on the thermal manifold breaks down for a general uncharged black hole at its Hawking temperature}. Despite several less rigorous steps in the above derivation\footnote{We did not explicitly construct the Zamolodchikov metric nor the three-point functions. The only thing we need from these however is that they are finite (and non-zero), which is something we did discuss here.}, we believe this conclusion is difficult to avoid. In any case, the one-loop result already shows that problems arise and this feature is apparently \emph{not} solved by summing higher genus contributions.\\

\noindent Other relevant work concerning the Hagedorn transition at finite string coupling (but using holographic methods instead) can be found in \cite{Liu:2004vy}\cite{AlvarezGaume:2005fv}\cite{Aharony:2003sx}. \\

\begin{figure}[h]
\centering
\includegraphics[width=0.5\linewidth]{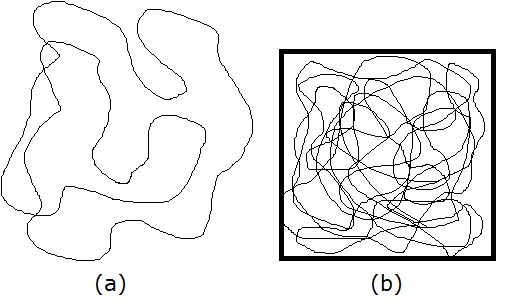}
\caption{(a) Random walk in an uncompact space. The random walk is not densily packed as it can safely avoid an unconstrained number of self-crossings. (b) In a compact space, the random walk is forced to pass through the same point over and over. This causes the density of string to diverge which manifests itself in an infinite free energy density.}
\label{cont}
\end{figure}

\noindent If $g_s$ is strictly zero however, all higher genus amplitudes vanish and Susskind's (non-interacting) long string picture is valid. As soon as any form of interaction is allowed ($g_s$ not strictly zero, but arbitrarily small), the higher genus amplitudes do not vanish and a resummation is necessary. This is in agreement with Susskind's prediction that interactions, if present, will \emph{always} become important close to the event horizon. The fact that setting $g_s$ to zero gives an infinite free energy density does not seem so strange: the string becomes arbitrarily long and intersects itself numerous times effectively giving an infinite density of string everywhere. This is intuitively obvious: non-compact dimensions are always larger than the string itself thus reducing the `chance' of a piece of string returning to the same location. For compact spaces, the string feels the boundaries and is forced to pass through the same location over and over again \cite{Barbon:2004dd}. A cartoon is given in figure \ref{cont}.
A related point was also made by Susskind: he argues that higher order interactions should become important as $g_s^2 \rho \ell_s^{d-2}$ becomes of order 1, with $\rho$ the number of string crossings per unit horizon area and $d$ the spacetime dimensionality. This argument was made in a single-string (microcanonical) picture. In our case, the condition (for the most dominating contribution; this is the worst case scenario) is instead $g_s^2\frac{1}{(\beta-\beta_H)^3} \ell_s^{3}$ of order 1. This suggests some proportionality:
\begin{equation}
\rho \propto \frac{\ell_s^{5-d}}{(\beta-\beta_H)^{3}}
\end{equation}
and equating the temperature to the Hawking temperature hence immediately leads to an infinite number of string crossings $\rho$ per unit horizon area. As long as $g_s \neq 0$, higher order terms in the perturbation series are not negligible.\\

\noindent We conclude that as soon as $g_s$ is non-zero, one should not use the perturbation series anymore.

\section{Conclusion}
Building upon our earlier work \cite{Mertens:2013zya}, we provided some additional ideas on the thermal scalar near black hole horizons (effectively in Rindler space) and the role it plays in describing string thermodynamics. We emphasized the relevance of the highly excited string for (one-loop) black hole thermodynamics: there is some tension between solely considering thermodynamical properties of massless fields and the entire string. We provided an argument in favor of the lowest order in $\alpha'$ thermal scalar action for heterotic strings in Rindler space. We also provided a more elaborate comparison of the large $k$ limit of the cigar orbifold partition functions and the partition functions of the flat cones. We found detailed agreement. The twisted sectors do not have IR divergences which we used to compare even the numerical prefactors. A formula was written down that relates the 2d area with the level $k$ and the modular invariant regulator $\epsilon$ introduced in \cite{Giveon:2014hfa}. Such a procedure led to a deeper understanding of the link between both models: in particular, the disappearance of higher winding modes and the link between continuations in $N$ in both models was elaborated upon. Using these exact limits as an inspiration, we discussed the two different approaches that can be identified to study string thermodynamics in a space with a contractible thermal circle. We were led to the fact that the exact CFT results include interactions with an open string gas on the horizon and that the free closed string trace is encapsulated in the path integral approach with boundary conditions with fixed wrapping numbers as we studied in the near-Hagedorn regime in \cite{theory}. The latter path integral can be directly related to such a path integral in the fundamental domain and hence is modular invariant as well. Both approaches lead to different physics: the full genus 1 result is to be combined with the higher genera to give the full thermal free energy. It is also this one that should be used to indicate backreaction of the thermal gas on the background. The fixed wrapping number path integral on the other hand describes manifestly non-interacting closed string propagation and hence this should correspond to the free-field trace. Next, we gave some additional remarks on the critical $k$ value for the $SL(2,\mathbb{R})/U(1)$ cigar from the thermal scalar perspective. Finally we analyzed the contributions to thermodynamics from higher genera. We concluded that for each fixed genus, the random walk can split with 3-point bifurcations (the number of such events determines the genus). The full perturbation series however needs to be resummed and the genus expansion itself appears to break down in this case. \\

\noindent As a general conclusion, it seems clear that the topic of string thermodynamics in a space with a horizon is still far from understood completely, but it is our hope that we have proposed some suggestive ideas (supported by computations) that would enable a full understanding of this subject in the (hopefully near) future.

\section*{Acknowledgements}
It is a pleasure to thank David Dudal for several valuable discussions. TM thanks the UGent Special Research Fund for financial support. The work of VIZ was partially supported by the RFBR grant 14-02-01185.

\appendix

\section{Spectrum on the $SL(2,\mathbb{R})/U(1)$ cigar CFT}
\label{cigspectr}
In this appendix, we write down the spectrum of strings on the $SL(2,\mathbb{R})/U(1)$ cigar CFT. The winding $w$ around the cigar and discrete momentum $n$ are combined into two combinations
\begin{equation}
\label{B8}
m = \frac{n+kw}{2}, \quad \bar{m} = \frac{-n+kw}{2}, \quad n,w \in \mathbb{Z}.
\end{equation}
The quantum number $j$ is proportional to the radial momentum and is given by
\begin{align}
j &= -\frac{1}{2}+is, \quad s\in\mathbb{R} , \quad \text{continuous representations}, \\
j &= M - l , \quad l=1,2,...,\quad \text{discrete representations},
\end{align}
where $M = \text{min}(m,\bar{m})$ with $m,\bar{m} > 1/2$ \cite{Aharony:2004xn}. For the discrete representations, $j$ has the following unitarity constraints:
\begin{align}
\label{unibos}
-\frac{1}{2} &< j < \frac{k-3}{2}, \quad \text{bosonic},\\
-\frac{1}{2} &< j < \frac{k-1}{2}, \quad \text{type II}.
\end{align}
The conformal weights of the states are given by
\begin{align}
h &= \frac{m^2}{k}-\frac{j(j+1)}{k-2}, \quad \bar{h} = \frac{\bar{m}^2}{k}-\frac{j(j+1)}{k-2} , \quad \text{bosonic},\\
h &= \frac{m^2}{k}-\frac{j(j+1)}{k}, \quad \bar{h} = \frac{\bar{m}^2}{k}-\frac{j(j+1)}{k} , \quad \text{type II}.
\end{align}

\section{Proof of the McClain-Roth-O'Brien-Tan theorem from the torus path integral}
\label{MROB}
In this appendix, we will be more explicit in the general proof of this theorem \cite{McClain:1986id}\cite{O'Brien:1987pn}. The proof splits in two parts: first the modular transformation properties of the two wrapping numbers is distilled. Then in the second part it is shown that these are sufficient to perform the extension to the strip domain while dropping one wrapping number. 
\subsection{Part 1: Modular transformations of the wrapping numbers}
Assume the general path integral on the fundamental domain has a periodic coordinate $X \sim X + 2\pi R$. Then this path integral is considered with the set of boundary conditions
\begin{align}
X(\sigma^1 + 2\pi , \sigma^2) &= X(\sigma^1,\sigma^2) + 2\pi w R, \\
X(\sigma^1 + 2\pi\tau_1 , \sigma^2+2\pi\tau_2) &= X(\sigma^1,\sigma^2) + 2\pi m R,
\end{align}
with a flat worldsheet metric in the action. This form is particularly suited to our task since all dependence on $\tau$ and on the wrapping numbers is fully extracted from the action into the boundary conditions. \\
The entire path integral is modular invariant and the modular group $PSL(2,\mathbb{Z})$ is generated by the $T$ and $S$ transformations. Consider first the $T$ transformation $\tau \to \tau +1$. The second boundary condition then gets translated into
\begin{align}
X(\sigma^1 + 2\pi(\tau_1 +1), \sigma^2+2\pi\tau_2) = X(\sigma^1 + 2\pi\tau_1 , \sigma^2+2\pi\tau_2) + 2\pi w R = X(\sigma^1,\sigma^2) + 2\pi m R,
\end{align}
yielding
\begin{align}
X(\sigma^1 + 2\pi\tau_1, \sigma^2+2\pi\tau_2) =  X(\sigma^1,\sigma^2) + 2\pi (m-w) R.
\end{align}
Hence upon making the replacement $m \to m+w$ (or $m'=m-w$), the invariance is established. \\

\begin{figure}[h]
\centering
\includegraphics[width=0.25\textwidth]{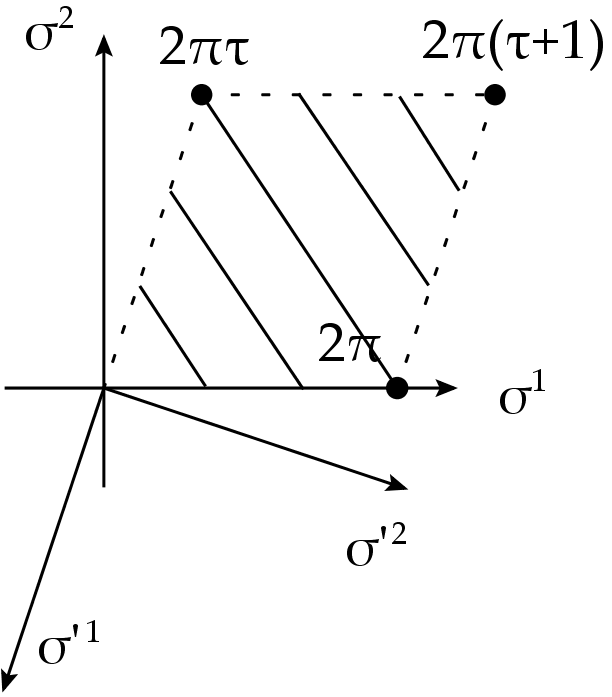}
\caption{Worldsheet coordinates and torus with modulus $\tau$. The new (primed) coordinates are also shown.}
\label{plane}
\end{figure}

\noindent Secondly we consider the $S$-transformation $\tau \to -1/\tau$. To start with, we first make a scaling coordinate transformation on the worldsheet as
\begin{equation}
z' = -\frac{z}{\tau}, \quad z = \sigma^1 + i \sigma^2, \quad z' = \sigma'^1 + i \sigma'^2.
\end{equation}
As long as the non-linear sigma model satisfies the Einstein equations, it is conformal and the above scaling symmetry has no effect on the action itself (it gets transformed into the same action with primed coordinates). The boundary conditions however are not the same anymore (corresponding to the fact that these transformations are not in the CKG of the torus). Explicitly, we obtain
\begin{equation}
\sigma'^1 = - \frac{\sigma^1\tau_1 + \sigma^2\tau_2}{\left|\tau\right|^2}, \quad \sigma'^2 = - \frac{-\sigma^1\tau_2 + \sigma^2\tau_1}{\left|\tau\right|^2}.
\end{equation}
These new coordinates are shown in figure \ref{plane}.
In these new worldsheet coordinates, the two torus cycles are described as
\begin{align}
(\sigma^1 , \sigma^2) \sim (\sigma^1 + 2\pi, \sigma_2) &\to (\sigma'^1, \sigma'^2) \sim \left(\sigma'^1 - \frac{2\pi \tau_1}{\left|\tau\right|^2}, \sigma'^2 + \frac{2\pi \tau_2}{\left|\tau\right|^2}\right), \\
(\sigma^1 , \sigma^2) \sim (\sigma^1 + 2\pi \tau_1, \sigma_2 + 2\pi\tau_2) &\to (\sigma'^1, \sigma'^2) \sim (\sigma'^1 - 2\pi , \sigma'^2).
\end{align}
Up to this point, only passive transformations were used. Next, we do the $S$-transformation, which yields the final transformed boundary conditions:
\begin{align}
X(\sigma'^1 + 2\pi \tau_1, \sigma'^2 + 2\pi\tau_2) &= X(\sigma'^1,\sigma'^2) + 2\pi w R, \\
X(\sigma'^1 - 2\pi, \sigma'^2) &= X(\sigma'^1,\sigma'^2) + 2\pi m R.
\end{align}
The substitution
\begin{equation}
w \to m ,\quad m\to -w,
\end{equation}
or ($w'=-m$ and $m'=w$) then shows that the $S$-transformation is a symmetry. \\
To summarize, the following modular transformation properties are found:
\begin{align}
T &: \quad m \to m+w , \\
S &: \quad m \to -w, \quad w \to m.
\end{align}

\subsection{Part 2: $SL(2,\mathbb{Z})$ manipulations}
The second phase of the proof consists of showing that the above modular transformation properties of $m$ and $w$ are enough to transform the fundamental domain into the strip. This part actually goes identically the same as for flat space \cite{McClain:1986id}\cite{O'Brien:1987pn}. Because of this reason, we will go rather quickly through the necessary steps. The goal is clear: find out what modular transformation on $\tau$ needs to be done to ensure that the ($m$, $w$) doublet gets transformed into ($m$, $0$) and at the same time build up the remaining regions of the strip. \\
Firstly, it is known that the set of modular transformations on $\tau$ (with composition of transformations) is isomorphic (as a group) to the set of $PSL(2,\mathbb{Z})$ matrices, equiped with matrix multiplication. A modular transformation and a $SL(2,\mathbb{Z})$ matrix are identified as
\begin{equation}
\frac{a\tau + b}{c\tau +d} \quad \Leftrightarrow \quad 
\left[\begin{array}{cc}
a & b  \\
c & d \end{array}\right],
\end{equation}
and the $T$ and $S$ generators of this group are given by
\begin{equation}
T = \left[\begin{array}{cc}
1 & 1  \\
0 & 1 \end{array}\right], \quad S = \left[\begin{array}{cc}
0 & -1  \\
1 & 0 \end{array}\right].
\end{equation}
As is well-known, $c$ and $d$ should be relatively prime by the determinantal condition $ad-bc=1$. The quantum numbers $m$ and $w$ then form a doublet of this group action, where the column vector $\left[\begin{array}{c}
m   \\
w  \end{array}\right]$ transforms under $PSL(2,\mathbb{Z})$ by left multiplication. The $SL(2,\mathbb{Z})$ element that we seek is hence of the form
\begin{equation}
\left[\begin{array}{cc}
a & b  \\
c & d \end{array}\right]\left[\begin{array}{c}
m   \\
w  \end{array}\right] = \left[\begin{array}{c}
X   \\
0  \end{array}\right],
\end{equation}
where the two bottom elements $c$ and $d$ are determined by $cm + dw = 0$ or $cm = -dw$. Let the gcd of $m$ and $w$ be $r$. Then the only possibility for relatively prime $c$ and $d$ is the solution $c=\frac{w}{r}$ and $d=-\frac{m}{r}$. \\
For fixed $c$ and $d$, the top elements $a$ and $b$ are restricted by the determinantal condition and one readily shows that the only freedom left is $(a,\, b) \to (a+\lambda c, \, c+\lambda d)$ for $\lambda \in \mathbb{Z}$. These elements can all be obtained from one such element $M$ by left multiplication by $T$ as $T^\lambda M$. Hence it is clear that only 1 value of $\lambda$ exists for which the resulting modulus is located in the strip region. Moreover, any two matrices that do not have the same $c$ and $d$ do not have any overlap in the strip region. It is also clear that the strip is built up entirely since $PSL(2,\mathbb{Z})$ fully generates the upper half plane from the fundamental domain. The $SL(2,\mathbb{Z})$ element is fully fixed by this and the transformed modulus is related to the initial one as
\begin{equation}
\tau' = \frac{a\tau+b}{c\tau+d}.
\end{equation}
The value $X$ equals $-r$ (due to the determinantal condition). \\

\noindent As a summary of these steps, in a schematic fashion, one manipulates the expression as
\begin{equation}
\label{scheme}
\sum_{m,w} \int_{\mathcal{F}(m,w)} \to \sum_{r=1}^{+\infty} \sum_{\left[c,d\right]=1}\int_{\mathcal{F}(m,w)} \to 2 \sum_{r=1}^{+\infty} \int_{E(-r,0)}
\end{equation}
and $r$ takes over the role of the single wrapping number in the strip. \\
Note further that the worldsheet transformation $z \to -z$ is equivalent to replacing $m\to-m$ and $w\to-w$. Doing this, one finds that $X=+r$ instead, showing the symmetry $r\to-r$ in the resulting path integral. We can hence double the range of $r$ which destroys an extra factor of 2 we created earlier (\ref{scheme}) by having both ($c$, $d$) and ($-c$, $-d$) map to the same $r$-number. \\

\noindent In all of these manipulations, the ($m=0,\, w=0$) sector transforms as a singlet under the modular group and this is hence not altered: no build-up of the strip domain is present for this state and in the end it is still integrated over the modular fundamental domain.

\section{Proof of the claims in section \ref{remarks}}
\label{proof}
Let us take the cigar metric and dilaton (for bosonic strings) as \cite{Dijkgraaf:1991ba}\cite{Giveon:2013ica}\cite{Mertens:2013zya}
\begin{align}
ds^2 &= \frac{\alpha'}{4}(k-2)\left[dr^2 + \frac{4}{\coth^2\left(\frac{r}{2}\right) - \frac{2}{k}}d\theta^2\right],\\
\Phi &= -\frac{1}{2}\ln\left(\frac{\sinh(r)}{2}\sqrt{\coth^2\left(\frac{r}{2}\right) - \frac{2}{k}}\right).
\end{align}
We rescale the coordinates as $\rho = \frac{\sqrt{\alpha'(k-2)}}{2}r$ and $\theta_{new} = \sqrt{\alpha'(k-2)}\theta_{old}$, which gives
\begin{align}
ds^2 &= \left[dr^2 + \frac{1}{\coth^2\left(\frac{\rho}{\sqrt{\alpha'(k-2)}}\right) - \frac{2}{k}}d\theta^2\right],\\
\Phi &= -\frac{1}{2}\ln\left(\frac{\sinh\left(\frac{2}{\sqrt{\alpha'(k-2)}}\rho\right)}{2}\sqrt{\coth^2\left(\frac{\rho}{\sqrt{\alpha'(k-2)}}\right) - \frac{2}{k}}\right).
\end{align}
Taking the coordinates very large $\rho \gg \sqrt{\alpha'(k-2)}$, we arrive at a linear dilaton background (as expected), but with an unconventional coordinate scaling:
\begin{align}
ds^2 &= \left[d\rho^2 + \frac{1}{1 - \frac{2}{k}}d\theta^2\right],\\
\Phi &= \Phi_0 -\frac{1}{\sqrt{\alpha'(k-2)}}\rho.
\end{align}
We see now that a further rescaling as $\theta_{new} = \sqrt{\frac{k}{k-2}}\theta_{old}$ yields the standard linear dilaton background.
The Hawking temperature in the new $\theta$ coordinate is $\beta_{\text{Hawking}} = 2\pi\sqrt{\alpha'k}$. The linear dilaton background has a Hagedorn temperature equal to
\begin{equation}
\beta_{H}^2 = 4\pi^2\alpha'\left(4-\frac{1}{k-2}\right).
\end{equation}
The value of $k$ where the asymptotic Hagedorn temperature (being the linear dilaton case) and the Hawking temperature are equal is at $k=3$. Analogously (but slightly easier) one can perform the computation for the type II superstring and one finds $k=1$ as the critical value.\\

\noindent Let us look at the cigar spectrum (written down in appendix \ref{cigspectr}) and determine when $w=1$ states drop from the spectrum. 
We focus on $n\geq 0$ states (the other case is actually the same). For a discrete state we have that
\begin{equation}
j = \frac{n+kw}{2} - l , \quad l = 1,2,\hdots
\end{equation}
The bosonic unitarity constraints reduce to (for $w=1$)
\begin{align}
l &> \frac{n+3}{2}, \\
l &< \frac{n+k+1}{2}.
\end{align}
Clearly the value of $n$ is irrelevant for the number of allowed $l$-values and only for $k > 3$, we can have states satisfying these constraints.
Moreover, at precisely $k=3$, the constraints 
\begin{align}
l &> \frac{n+3w}{2}, \\
l &< \frac{n+3w+1}{2},
\end{align}
cannot be satisfied for \emph{any} $n$ and $w$, implying that there are no discrete states at all. For lower values of $k$ some discrete states may reappear into the spectrum. One can also see this in the thermal partition function \cite{Hanany:2002ev}, where these unitarity constraints are represented as a contour-shift that either encircles poles or not.
For the type II superstring, the first constraint is replaced by $l > \frac{n+1}{2}$ leading to $k=1$ as the critical value.\\

\noindent One can also find this critical value of $k$ by looking at the asymptotics (large $\rho$) of the wavefunctions (imposing normalizability). Setting $\alpha'=2$, the measure factor $\sqrt{G}e^{-2\Phi}$ is proportional to 
\begin{equation}
e^{\sqrt{2/(k-2)}\rho}
\end{equation}
and the states have asymptotics 
\begin{equation}
\label{stasympt}
\psi \propto e^{-\sqrt{2/(k-2)}(j+1)\rho},
\end{equation}
which leads to the same value $k=3$ (or $k=1$ for type II strings). \\

\noindent A final characterization of this value is found by looking at the continuum. Assuming that the continuous quantum number $s$ does not give a contributing $\tau_2$-dependent exponential correction (as we have seen happens in fact for the linear dilaton background discussed elsewhere \cite{Mertens:2014cia}), we find that the continuum state becomes marginal when\footnote{In principle one should integrate over the $s$-quantum number. The density of states of this model and that of the $AdS_3$ WZW model are given by the same expression, and we have shown in \cite{Mertens:2014nca} that the integration over $s$ can not yield a correction to the critical temperature.}
\begin{equation}
\frac{1}{4(k-2)} + \frac{k}{4} = 1 \Leftrightarrow k=3
\end{equation}
and analogously for the type II superstring.\\
We remark that for both $k$ larger and smaller than this critical value, this state is non-tachyonic, it can only become marginal when this critical value of $k$ is reached.\\
Note also that this state is the thermal scalar of the asymptotic linear dilaton background. Indeed, allowing $\beta$ to vary, we would write\footnote{From the cigar perspective, this off-shell generalization can be found by first considering conical orbifold models. This would cause the replacement $\frac{kw^2}{^4} \to \frac{kw^2}{4N^2}$ in the conformal weights. Then we reinterpret $\frac{1}{N}$ as $\frac{\beta}{\beta_{\text{Hawking}}}$ to arrive at the correct result.}
\begin{equation}
\frac{1}{4(k-2)} + \frac{k}{4}\frac{\beta^2}{4\pi^2k\alpha'} = 1 \,\, \Leftrightarrow \,\, \beta_H = 2\pi\sqrt{\left(4-\frac{1}{k-2}\right)\alpha'},
\end{equation}
or for type II strings:
\begin{equation}
\frac{1}{4k} + \frac{k}{4}\frac{\beta^2}{4\pi^2k\alpha'} = \frac{1}{2} \,\, \Leftrightarrow \,\, \beta_H = 2\pi\sqrt{\left(2-\frac{1}{k}\right)\alpha'},
\end{equation}
which are indeed the correct Hagedorn temperatures of the linear dilaton background. In particular for only flat extra dimensions, we can relate this to the non-critical flat spaces upon setting $Q^2 = \frac{4}{(k-2)\alpha'} = 4\frac{26-D}{6\alpha'}$,\footnote{For type II superstrings, we have instead $Q^2 = \frac{4}{k\alpha'} = \frac{10-D}{\alpha'}.$} where the parameter $Q$ is conventionally defined as $\Phi = \frac{Q}{2}X^d$ with $d$ the linear dilaton direction.

\section{Discussion on non-normalizable versus normalizable vertex operators}
\label{nnversusn}
The computations given in \cite{Giveon:1999px}\cite{Giveon:1999tq} and \cite{Aharony:2004xn} focus on non-normalizable observables and these give a series of poles for the $n$-point functions. One should distill the residue of these poles to obtain the S-matrix elements. These then correspond to $n$-point functions of the associated normalizable operators and these should be finite. This solves an initial worry one might have in that the Zamolodchikov metric might diverge or the 3-point functions might vanish. As an extra reassurance, the four-point amplitude for non-normalizable observables for the $AdS_3$ model was computed in \cite{Maldacena:2001km} where an intermediate virtual discrete state was found indeed, suggesting the inverse Zamolodchikov metric for such states is non-zero. These correlation functions in $AdS_3$ are directly related to those of the cigar CFT and again one should zoom in on the LSZ residue instead to obtain the required stringy amplitudes obtained by normalizable vertex operators. \\
The link between the normalizable and non-normalizable vertex operators is
\begin{equation}
\mathcal{O}_{\text{non-norm}} = \frac{1}{\text{pole}}\mathcal{O}_{\text{norm}}.
\end{equation}
In more detail, for large $\phi$ the vertex operators behave as \cite{Aharony:2003vk}:\footnote{In this expression, $Y$ denotes a free boson and $\phi$ is a coordinate coming from the $AdS_3$ ancestor of this vertex operator. In the notation given in this paper, $\phi=\rho$, the rescaled radial coordinate introduced in section \ref{recap}.}
\begin{equation}
V_{j,m,\bar{m}} \sim e^{iQmY}\left(e^{Qj\phi} + \frac{1}{\text{pole}} e^{-Q(j+1)\phi} + \hdots\right).
\end{equation}
Then multiplying by the pole contribution distills purely the normalizable part of the vertex operator. Indeed, it is this asymptotic behavior that we used in constructing the normalizable string fluctuations for Euclidean Rindler space in \cite{Mertens:2013zya}. The reader should compare this to equation (\ref{stasympt}). Moreover, the pole spectrum of the non-normalizable 2-point functions, originally constructed to be the spectrum of states in the dual LST \cite{Aharony:2004xn}, is also the discrete spectrum of the cigar CFT.\footnote{Note that in \cite{Hanany:2002ev}, the cigar partition function was rewritten in CFT language and the spectrum was distilled. The precise discrete spectrum of appendix \ref{cigspectr} was not found though it seems obvious that a more careful study of their argument would indeed reproduce the spectrum. Also, interestingly, we saw in \cite{Mertens:2014nca} that the spectrum of discrete states in thermal $AdS_3$ orbifolds actually also shows precisely this same structure.} The discrete part of both Hilbert spaces is then indeed the same. \\
Since the 2-point function of non-normalizable vertex operators has only a single pole, we immediately deduce that the 2-point function of normalizable operators (the scattering amplitude) vanishes, just like in flat space. Schematically,
\begin{equation}
\left\langle \mathcal{O}_{\text{norm}} \mathcal{O}_{\text{norm}} \right\rangle = \text{pole}^2\left\langle \mathcal{O}_{\text{non-norm}} \mathcal{O}_{\text{non-norm}} \right\rangle = 0. 
\end{equation}

\end{document}